# Inner-shell photodetachment from $Si^-$ negative ion: strong effect of many-electron correlations

G Schrange-Kashenock

*St.Petersburg Polytechnical University, St.Petersburg 192251, Russia*
e-mail: gkashenock@yahoo.de

**Abstract.** The first theoretical investigation on the inner-shell single-photodetachment from the $Si^-$ ($1s^22s^22p^63s^23p^3\ {}^4S^o$) negative ion is presented. The partial and total cross sections, the photoelectron phaseshifts, and the parameters of angular anisotropy are calculated in the framework of Many-Body Theory for L-shell photodetachment from $Si^-$ ion in the experimentally accessible range of photon energies (7.5-14 *Ry*). Comparison is made between the calculations of the response of the ionic many-electron system $Si^-$ to an electromagnetic field at the different levels of approximation: the "frozen-field" Random Phase Approximation with Exchange (RPAE), and the static relaxation approximation. The optimal analysis is made when the dynamic relaxation and polarization are included within the Dyson Equation Method (DEM) simultaneously with the RPAE corrections (RPAE&DEM approach). It is predicted that the photoexcitation to a resonance state of complex "shape-Feschbach" nature in the open p-shell reveals itself as a prominent resonance structure in the photodetachment cross sections in the energy range of the 2s and 2p inner shell thresholds similar to that in 1s inner-shell photodetachment from $C^-$ (*J.Phys.B* 2006 **39** 1379). The photodetachment dynamical characteristics clearly demonstrate the significance of all the considered many-electron correlations within the RPAE&DEM approach, however the total photodetachment cross section is dominated by a strong resonance peak just after the 2s threshold. Dynamical relaxation (screening) is identified as a decisive factor in the formation of this resonance.

## 1. Introduction

The photodetachment from deep inner shells of negative ions stands out as an extremely sensitive probe and theoretical test-bed for important effects of electron-electron interaction because of the weak coupling between photon and target electrons. Many-body effects play a pronounced role here not only between the inner electrons but also between the outer-shell electrons and outgoing electron. Many-electron correlations are of prominent importance in photodetachment from negative ions of any atomic systems, especially with resonance processes inherent. In the recent years, significant progress has been achieved using the merged ion beam-photon beam technique (Kjeldsen 2006), which fact is demonstrated, in particularity, by inner-shell photodetachment experiments on $C^-$ (Walter *et al* 2006) and $B^-$ (Berrah *et al* 2007) at the Advanced Light Source beamline 10.01.1. New possibilities for precise measurements of transition energies have been provided by the high photon-energy resolution available at third generation synchrotron light sources (Müller 2015). The latest tunable infrared laser valence-shell photodetachment spectroscopy experiments show that the negative ion $La^-$ is a very promising negative ion for laser-cooling applications (Walter *et al* 2014). The resonant photodetachment from the valence shells of cerium $Ce^-$ negative ion has also been investigated using this experimental method (Walter *et al* 2011). Using Advanced Light Source photons, the experimental photodetachment cross sections have also been obtained for the fullerene negative ion $C_{60}^-$ (Bilodeau *et al* 2013). Many-Body Theory methods are a powerful tool for the theoretical study of systems with inherent strong correlation interaction and show promise of precision calculations of structures and photodetachment/photoionisation characteristics for complex, strongly correlated atoms/ions, clusters, and fullerenes. In the present investigation the case of an atomic strongly correlated system – the open-shell negative silicon ion – will be considered.

## 2. Problem and approach

In the present report theoretical results on the inner-shell photodetachment from the $Si^-$ ($1s^22s^22p^63s^23p^3$ $^4S^o$) negative ion will be presented. One can expect that the possibility of a photoexcitation to the ion state $Si^{-*}$ ($1s^22s2p^63s^23p^4$ $^4P$) reveals itself as a resonance structure in photodetachment cross sections in the energy range of the 2s and 2p inner shells thresholds similar to that in the 1s inner-shell photodetachment from $C^-$. For the carbon ion, the strong near-threshold resonance has been predicted within the approach referred to as RPAE&DEM (Kashenock and Ivanov 2006) in a good agreement with the experiment (Walter *et al.* 2006) and other theoretical results (Gibson *et al.* 2003, Corgyca 2004, Zhou *et al.* 2003). The theoretical analysis reveals the complex, mixed ("shape-Feschbach" hybrid) nature of the resonance. The RPAE&DEM approach has been developed in a series of studies (Ivanov *et al.* 1996, Kashenock and Ivanov 1997, 1998, 2006) for the simultaneous inclusion of the dynamic polarization potential generated in the system of "a neutral atomic core and an extra electron" and the dynamic relaxation (screening) and corrections within the framework of the Random-Phase Approximation with Exchange (RPAE). It uses the Hartree-Fock approximation (HF) as a zero-order approximation. Many-electron correlations are incorporated within the RPAE (to describe the dynamic collective response of an atomic system to the external field – interchannel and intrachannel many-electron correlations) and the Dyson Equation (Dyson 1949) Method (DEM, to correct electron behaviour due to dynamical polarization and relaxation effects). The collective character, the importance of electron correlations, for a response to the external electromagnetic field in the few-electron strongly correlated $C^-$ target is clearly demonstrated in the work by Kashenock and Ivanov (2006), with the dynamical relaxation of the core being the most pronounced of the correlation effects. The silicon negative ion, that has the same open p-shell structure, could be an even more intriguing object for many-body-theoretical study.

The ground state of the silicon negative ion within the Spin-Polarized Hartree-Fock (SPHF) approximation reads as $Si^-$ $1s{\uparrow}1s{\downarrow}2s{\uparrow}2s{\downarrow}2p^3{\uparrow}2p^3{\downarrow}3s{\uparrow}3s{\downarrow}3p^3{\uparrow}$ ($^4S$), where we consider subshells with spin up (↑) and spin down (↓) as closed. We use the spin-polarized (SP) approximations for the RPAE (Amusia 1990) as well as the SP generalization of the RPAE&DEM method. The exchange interaction occurs only between electrons with the same spin. This results in the energy "splitting" of "up" and "down" spin-polarized subshells. The system $Si^-$ is more complex compared to $C^-$ where the resonance is primarily formed by the one-channel contribution. Here we need to consider the partial cross sections for four close spin-polarized inner subshells (six phototransitions):

$Si^-$ ... $2p^3{\uparrow}2p^3{\downarrow}\ldots3p^3{\uparrow}$ ($^4S$) + $\omega \rightarrow$ Si ... $2p^2{\uparrow}2p^3{\downarrow}\ldots3p^3{\uparrow}$ ($^3P$) + $\varepsilon s{\uparrow}$ and $\varepsilon d{\uparrow}$ ;

$Si^-$ ... $2p^3{\uparrow}2p^3{\downarrow}\ldots3p^3{\uparrow}$ ($^4S$) + $\omega \rightarrow$ Si ... $2p^3{\uparrow}2p^2{\downarrow}\ldots3p^3{\uparrow}$ ($^5P$) + $\varepsilon s{\downarrow}$ and $\varepsilon d{\downarrow}$ ;

$$Si^- \ldots 2s\uparrow 2s\downarrow \ldots 3p^3\uparrow\ (^4S) + \omega \rightarrow Si \ldots 2s\downarrow \ldots 3p^3\uparrow\ (^3S) + \varepsilon p\uparrow; \quad (1)$$
$$Si^- \ldots 2s\uparrow 2s\downarrow \ldots 3p^3\uparrow\ (^4S) + \omega \rightarrow Si \ldots 2s\uparrow \ldots 3p^3\uparrow\ (^5S) + n,\varepsilon p\downarrow;$$

or, using the simpler specification, we consider six phototransitions: $2p\uparrow\rightarrow\varepsilon s\uparrow$, $2p\uparrow\rightarrow\ \varepsilon d\uparrow$, $2p\downarrow\rightarrow\varepsilon s\downarrow$, $2p\downarrow\rightarrow\ \varepsilon d\downarrow$, $2s\uparrow\rightarrow\varepsilon p\uparrow$, $2s\downarrow\rightarrow\ n,\varepsilon p\downarrow$. For the last phototransition in (1) we have emphasized the existence of photoexcitation to the $3p\downarrow$ (n=3) discrete (or quasi-bound) state in the half-filled outer p-shell. This channel is expected to be a "resonance channel" as we have seen in the case of "$1s2s^22p^4$" resonance in $C^-$. The resonance channel for $Si^-$ inner-shell photodetachment is open at the $2s\downarrow$ threshold ($E_{2s\downarrow}^{SPHF}$=-11.784 *Ry*) in the vicinity of the thresholds of the others inner-shells photodetachment channels: $E_{2s\uparrow}^{SPHF}$=-11.806 *Ry*, $E_{2p\uparrow}^{SPHF}$=-8.020 *Ry*, $E_{2p\downarrow}^{SPHF}$=-7.973 *Ry*. In contrast to the carbon negative ion, the interchannel interaction for these energy-close subshells becomes very important and should be accounted for within the RPAE. The background cross section corresponding to photodetachment from the outermost subshells of $Si^-$ at the energy range of interest is negligible. These channels open at the much smaller energy (the theoretical threshold energy is $E_{3p\uparrow}^{SPHF}$=-0.124 *Ry* and the experimental electron affinity is $E_A^{exp}$(Si)=-0.1018 *Ry* (Kasdan *et al.* 1975)) and the contribution of the valence shells and their coupling with the channels (1) in the further calculations will not be included.

In this work we perform the calculation of the response of the ionic many-electron system $Si^-$ to an electromagnetic field at the different levels of approximation: the "frozen-field" RPAE, the static relaxation approximation (Generalized RPAE), and also in the framework of the RPAE&DEM approach when the dynamic relaxation and polarization can be included simultaneously with the RPAE corrections.

### 3. Calculations within the "frozen-field" and static relaxation approximation

Fig.1 presents the results of calculations in the "frozen-field" spin-polarized RPAE (Amusia 1990) when the photoelectron wavefunctions are obtained with the same SPHF Hamiltonian as for the initial $Si^-$ $1s\uparrow 1s\downarrow 2s\uparrow 2s\downarrow 2p^3\uparrow 2p^3\downarrow 3s\uparrow 3s\downarrow 3p^3\uparrow$ configuration but with a hole in the corresponding subshell $2s\uparrow$, $2s\downarrow$, $2p\uparrow$ or $2p\downarrow$. In this approximation, the "$3p\downarrow$" resonance state is a real bound state of an electron in the field of the $Si^-$ ion's configuration with a hole $2s\downarrow$ (the neutral core). It should be noted that the existence of discrete excited photoelectron states is not typical for negative ions. However, within the "frozen-field" approximation for ions with a half-filled outer p-shell, like the present case of $Si^-$ or $C^-$ considered before (Kashenock and Ivanov 2006), the $p\downarrow$ photoelectron finds itself inside the very diffuse p-cloud where the attractive Coulomb potential of the nucleus is not completely screened by the other electrons. Moreover, an extra electron interacts with a hole as with a positive particle – this interaction is accounted by calculating a photoelectron wave function in the way described above. As result, even in the one-particle approximation an extra $p\downarrow$ electron can be bound by the neutral core to a state that lies below the threshold. One can see that the results for partial cross sections in the one-particle HF approximation and the RPAE are rather different due to the inclusion of electron correlations that correspond to the intrachannel interaction for 2p subshells and interchannel interaction between all six channels. The oscillator strength $f^{SPHF}_{2s\downarrow\to 3p\downarrow}$ = 0.0536 corresponding to the phototransition 2s↓→ 3p↓ reveals itself only as the interference of Fano-profiles in the partial cross sections of photodetachment from 2p subshells at the photon energies corresponding to the energy of transition into 3p↓ state ($E^{SPHF}_{3p\downarrow}$=-0.31256 *Ry*). Due to the interchannel interaction the part of the oscillator strength transferred to the other channels and $f^{RPAE}_{2s\downarrow\to 3p\downarrow}$ is 0.02925 in the length (***r***-) representation of the dipole operator, or 0.03329 in the velocity ($\nabla$ -) form. The partial cross section for the 2s↓→ εp↓ channel gives only a very weak background ≈ 0.2 *Mb* that even becomes less when the interactions with the other channels is allowed for.

To consider the other limit of approximation we use the Generalized RPAE (GRPAE, Amusia 1990) method. The photoelectron wavefunctions are calculated in the field of the completely rearranged spin-polarized Hartree-Fock configurations of the neutral silicon core without a detaching electron in the corresponding subshell. The results of partial cross sections calculations are presented in fig.2. This approach (and its relativistic analogue (Radojević *et al* 1989, Kutzner 2004)) includes a simplification that the core rearrangement due to a hole creation is taken into account instantly at wherever the photon is absorbed and the photoelectron starts to move. It corresponds to the *static relaxation* approximation. Until recently, without having the RPAE&DEM approach available, we had to use only this approach suggesting that the relaxation time is less than the electron escape time. It is worth stressing here that within the static approximation the role of relaxation effects often turns out to be overestimated, which distorts the near-threshold photodetachment picture. Our calculations for $Si^-$ show that at the GRPAE level of approximation a p-photoelectron cannot be bound into the 3p↓ state in the SPHF field of the rearranged Si atom without a 2s↓ electron. Calculated within the GRPAE, *i.e.* within the static relaxation approximation, the neutral atomic core is compact, the attractive potential of a nucleus for an extra electron is strongly shielded, the hole is not exists. However, quasi-bound state, the shape resonance, can be formed due to do possible for a photoelectron to occupy a "quasi-vacant" state in the open 3p-shell. According to our calculation, the p↓-wave phaseshift demonstrates the typical behavior for a shape resonance: $\delta^{GRPAE}_{1\downarrow}(\varepsilon)$ sharply increases from the value $\pi$ at zero photoelectron energy ε to 5.2 over a narrow photoelectron energy range (inset in fig.2). Note, that in the "frozen-field" approximation $\delta^{HF}_{1\downarrow}(\varepsilon) \to 2\pi,\ \varepsilon \to 0$ - see fig.5 below - that corresponds according to Levinson's theorem to an existence of two bound p↓ states. Correspondingly, in the partial 2s↓→ εp↓ cross section a resonant near-threshold peak appears (fig. 2), however it is not strong - the maximum value is only ≈ 2.5 *Mb* in the one-particle approximation. The value of the peak is 1*Mb* when the interchannel interaction is accounted for since its oscillator strength is transferred due to electron correlations, and the "window" resonance profile in the 2p partial cross sections is formed in the partial cross sections for the 2p subshells. The interference profiles arise just after the 2s↓ threshold. It is worthwhile to note that in the static relaxation approximation the weak shape resonance structure in the 2s↑→ εp↑ channel shown in fig.1 ("frozen field") disappears – the quasi bound "4p↑" state does not exists in this approximation of strong relaxation, when the electron cloud is not as diffuse as in the frozen-field of a neutral core (SPHF $Si^-$ with the hole 2s↑ ) and the attractive potential of a nuclei is completely screened. The total cross sections for $Si^-$ in the region of the 2s and 2p thresholds reveals no prominent resonance structure within the GRPAE or within the "frozen-field" the approximation (fig.3).

The existence of both limits of the "3p↓" state - as a bound state, or Feschbach resonance, in the "frozen-core" approximation and as a quasi-bound state, shape resonance, in the static relaxation approximation - allow us to suppose that the real situation is subtler and could be close to that in the $C^-$ inner-shell photodetachment. Due to strong electron correlations in many-electron system the resonance type could be considered as a mixed Feschbach-shape structure – we can expect to find the trace of this state in the photodetachment dynamical characteristics as a hybrid Fano-shape profile. To investigate the many-electron mechanism of forming the resonance near the 2s↓ threshold in detail we have used RPAE&DEM approach. For details of this approach, the reader is referred to the paper by Kashenock and Ivanov (2006) where the inner-shell photodetachment specifics of the approach are discussed or to the review for the RPAE&DEM theoretical formalism by Schrange-Kashenock (2015). Below only the essence of the method will be presented.

## 4 . RPAE&DEM formalism

When the system of "a neutral atomic core and an extra electron" is to be considered, the polarisation interaction and exchange interaction between an atom and the extra electron should be taken into account. Starting from the HF basis, the wavefunction $\psi_E(\vec{r})$ describing the motion of the electron with energy $E$ in the atomic field satisfies the Schrödinger-like Dyson equation:

$$\hat{H}^{(0)}\psi_E(\vec{r}\,) + \int \Sigma_E(\vec{r},\vec{r}\,')\psi_E(\vec{r}\,')d\vec{r}\,' = E\,\psi_E(\vec{r}\,)\,. \qquad (2)$$

Here $\hat{H}^{(0)}$ is the static Hartree-Fock Hamiltonian of the atom and $\Sigma_E(\vec{r},\vec{r}\,')$ is the energy-dependent non-local potential. This equation completely includes the correlational interaction of the extra electron with the atom. $\Sigma_E(\vec{r},\vec{r}\,')$ is equal to the self-energy part of the single-electron Green function of the atom, hence it can be presented as a diagrammatic expansion in powers of the interelectron correlational interaction. The dynamic non-local polarization potential may be built up *ab initio* within the Simplified RPAE (SRPAE, Chernysheva *et al* 1988). Starting from the HF basis $\Sigma$ is calculated in the second order of perturbation theory:

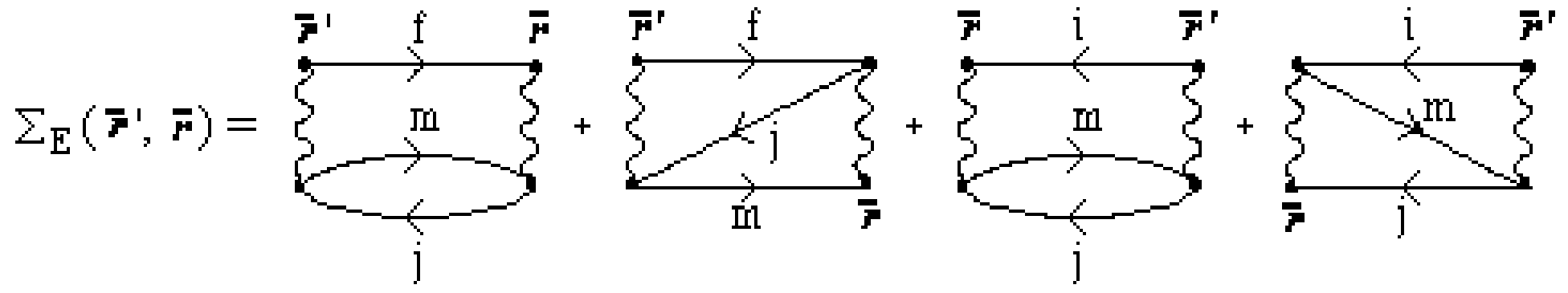

(3)

Hereafter, the Feynman-Goldstone diagram technique will be used (see *e.g.* Mattuck 1976, Amusia 1990) : the line with an arrow to the left (right) represents the hole (particle) which is below (above) the Fermi level; the wavy line is the Coulomb interaction. The complete orthonormal set of eigenfunctions $\varphi_\nu(\vec{r})$ which satisfy the equation $\hat{H}^{(0)}\varphi_\nu(\vec{r}) = \varepsilon_\nu \varphi_\nu(\vec{r})$ includes wavefunctions of discrete states ($\nu$ = *nl*: *j, i* in diagrams) and continuum states ($\nu$ = $\varepsilon l$: *f, m* in diagrams), which describe the electron of energy $\varepsilon$ and orbital momentum *l*, scattered by the Hartree-Fock potential.

We can construct now the reducible self energy part $\tilde{\Sigma}$ of the single-particle Green's function which is a solution of the following Dyson equation in a diagrammatic representation:

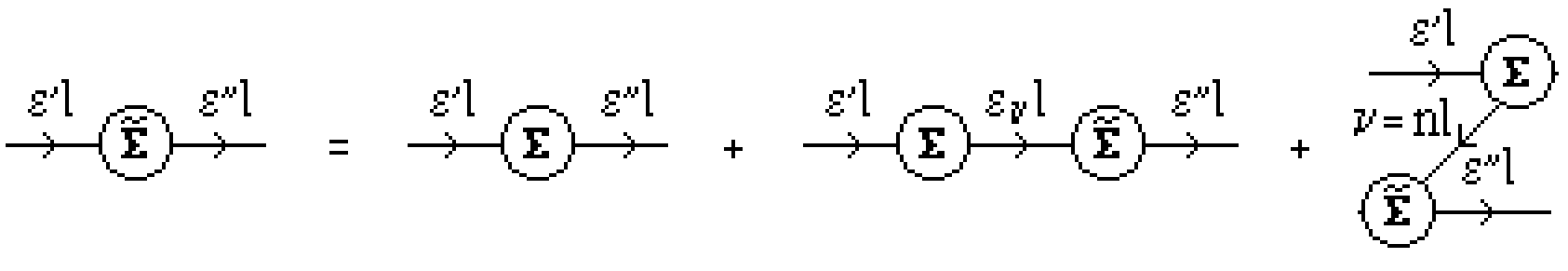

(4)

To incorporate the dynamic core-relaxation effects one further type of diagram comprising the mass operator $\Sigma$ contained in a kernel of the Dyson integral equation (2) should be included. The following equation

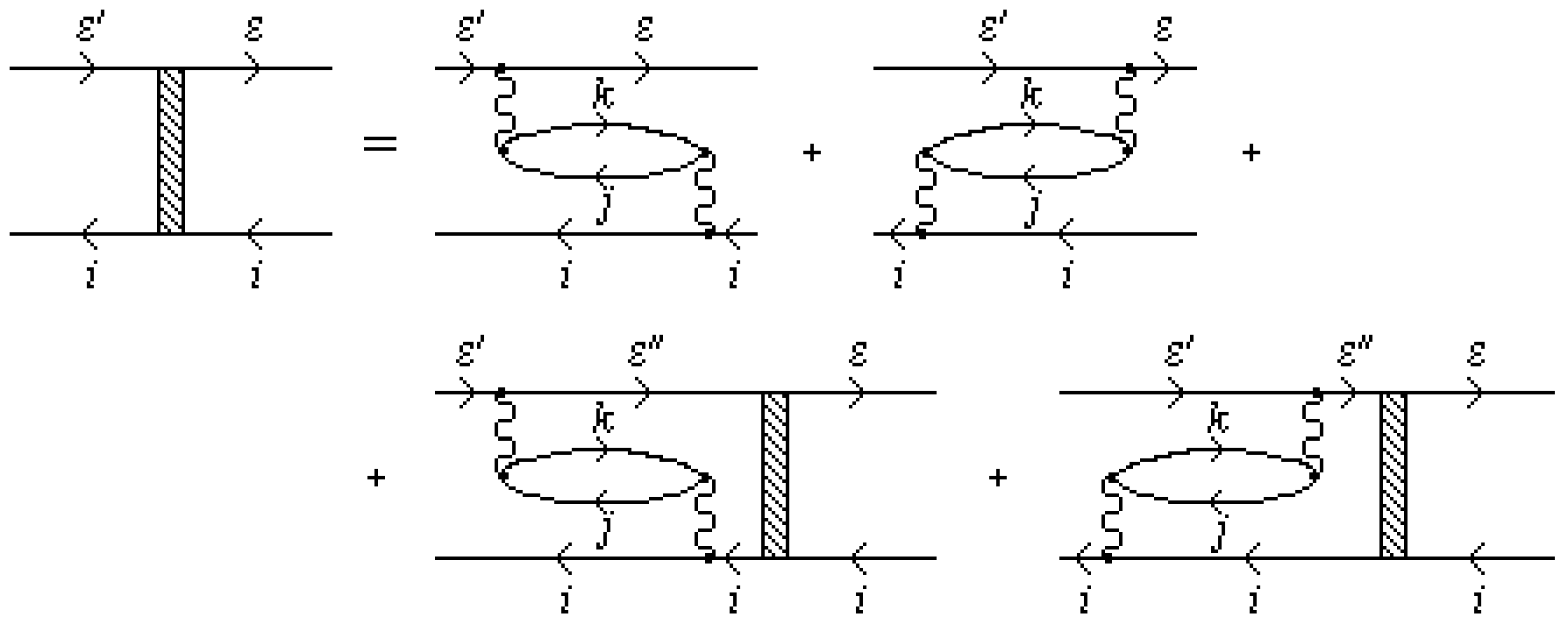

(5)

is solved and in doing so the screening interaction matrix elements $\langle \varepsilon | \tilde{\Gamma} | \varepsilon' \rangle$ are found.

To account for the influence of the atom-core dynamic polarization potential and effect of dynamical relaxation on an outgoing photoelectron the following effective dipole amplitudes are used

$$\langle \varepsilon | \hat{D}_\varepsilon | \tilde{i} \rangle = \frac{1}{1 + i\pi \langle \varepsilon | \hat{\tilde{\Sigma}}_\varepsilon | \varepsilon \rangle} \left( \langle \varepsilon | \hat{d} | \tilde{i} \rangle + vp \int_{\varepsilon'} \frac{\langle \varepsilon | \hat{\tilde{\Sigma}}_\varepsilon | \varepsilon' \rangle \langle \varepsilon' | \hat{d} | \tilde{i} \rangle}{\varepsilon - \varepsilon'} \right), \qquad (6)$$

where the reducible self-energy part $\tilde{\Sigma}$ consists of the dynamical relaxation $\hat{\Gamma}$ and polarization $\hat{\Sigma}$ parts,analytically:

$$\langle \varepsilon'' | \hat{\tilde{\Sigma}}_{\varepsilon_0} | \varepsilon' \rangle = \langle \varepsilon'' | \hat{\Sigma}_{\varepsilon_0} | \varepsilon' \rangle + vp \int_{\varepsilon_\nu} \frac{\langle \varepsilon'' | \hat{\tilde{\Sigma}}_{\varepsilon_0} | \varepsilon_\nu \rangle \langle \varepsilon_\nu | \hat{\Sigma}_{\varepsilon_0} | \varepsilon' \rangle}{\varepsilon_0 - \varepsilon_\nu} + \langle \varepsilon'' | \hat{\Gamma}_{\varepsilon_0} | \varepsilon' \rangle + vp \int_{\varepsilon_\nu} \frac{\langle \varepsilon'' | \hat{\tilde{\Sigma}}_{\varepsilon_0} | \varepsilon_\nu \rangle \langle \varepsilon_\nu | \hat{\Gamma}_{\varepsilon_0} | \varepsilon' \rangle}{\varepsilon_0 - \varepsilon_\nu} . \qquad (7)$$

Here $\hat{d}$ is the single-particle dipole operator in the length (r-) or velocity ($\nabla$-) form, and *vp* is the principal value symbol. Matrix elements are determined through the use of the HF wave functions $\varphi_\nu = | \nu \rangle$ and the ground state wavefunction is corrected within the DEM $| \tilde{i} \rangle$ by equation (2); we often drop some quantum numbers of the set determining the one-electron SPHF state $\nu = (n(\varepsilon), l, m, \mu)$ for the sake of brevity. Here $\varepsilon_0 = \omega + \varepsilon_i$ is an energy parameter, ω is a photon energy, $\varepsilon_i$ is the initial state energy. Throughout this paper $\int_{\varepsilon_\nu}$ denotes the summation and integration over the whole spectrum of ν states, occupied as well as excited. The matrix elements $\langle \varepsilon'' | \hat{\Gamma}_{\varepsilon_0} | \varepsilon' \rangle$ correspond to the following diagrams in the second order of perturbation theory:

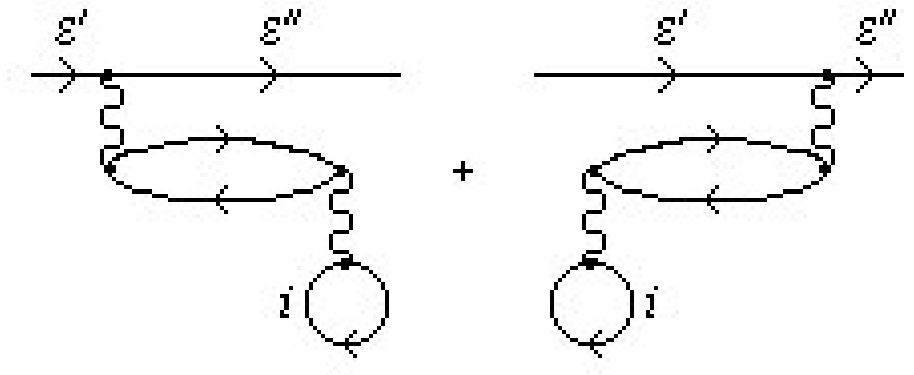

(8)

The effective dipole matrix elements (6) are calculated on the energy-shell, *i.e.* the state of energy ε is a real final state for the photodetachment process. In the work by Kashenock and Ivanov (1997) the general expression is evaluated for the effective dipole matrix elements of the energy-shell $\varepsilon \neq \varepsilon_0$, *i.e.* for phototransition to virtual states. Using the latter, one can solve the standard system of the RPAE equations and in so doing consider the intra- and interchannel interaction together with DEM electronic correlation corrections.

### 5. Calculation within the RPAE&DEM and conclusion

The one-particle approximation corresponds to that accepted in the "frozen-field" approximation. Our main result is that when the dynamical relaxation and polarization are included within the DEM even only for one channel, the resonance 2s↓→ εp↓ channel, the photodetachment pattern is changed dramatically. When the photodetachment amplitudes for the resonance 2s↓→ εp↓ channel are found from equation (6), where the reducible self-energy part $\tilde{\Sigma}$ consists of the dynamical relaxation (screening) $\hat{\Gamma}$ (8) and polarization $\hat{\Sigma}$ (3) parts, the 2s↓→ εp↓ partial cross section reveals a very strong, narrow near-threshold resonance (fig.4). In this one-channel consideration the most important diagrams are those corresponding to the infinite series from the diagrammatic (Feynman-Goldstone technique) equation:

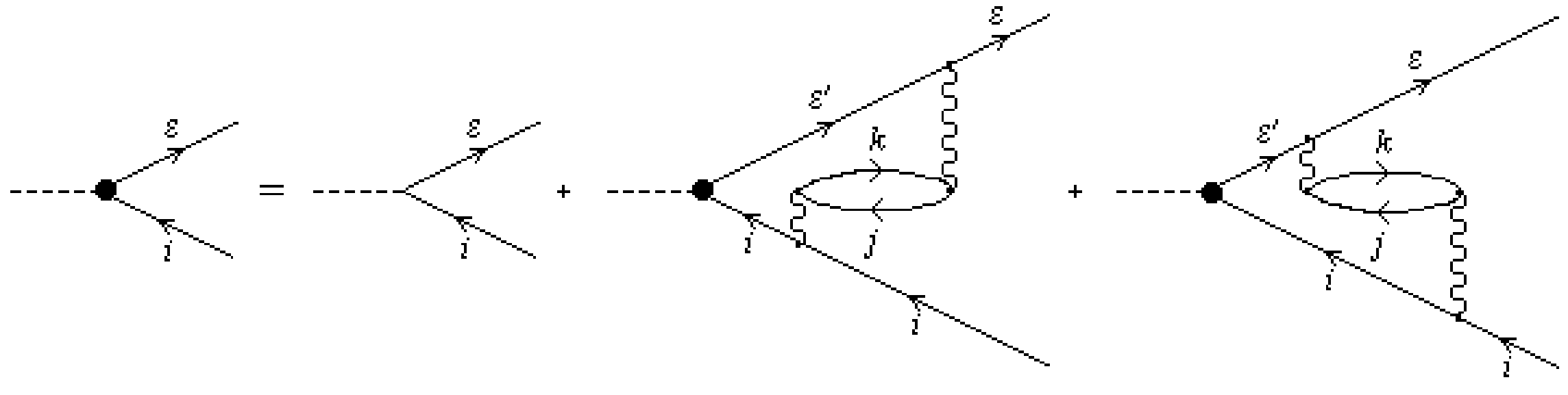

. (9)

Here, photoelectron $\varepsilon$ moves in the field of hole $i = 2s \downarrow$ which is strongly screened by the virtual excitations of electrons from the nearest subshells $j \to k$, the monopole excitations with $\varepsilon' =$ "3p↓" being the most pronounced. Photoelectron wavefunctions are calculated in the "frozen-core" field of an ion with a $i$ hole; the first term of equation (9) includes the hole field. The dipole matrix elements with allowance made for screening dynamic core relaxation (interaction with the photon - dashed line - denoted by closed circles), the new effective photodetachment amplitude, comprises a sequence of diagrams with an infinite number of screening loops. Within the DEM description of the relaxation process the interaction with a hole is weakened in comparison with the "frozen" field approximation. However, in contrast to the static approximation, the hole is still exists, only screened, and binding of an extra electron to a quasi-bound state in an open shell, a shape resonance, is very strong. The reaction of the core on removing of a photoelectron is described in our approach as a dynamical process - we can speak about *dynamical* relaxation. The present calculation has shown that this effect is the most important for the considered resonant photodetachment process.

In the frame of the RPAE&DEM approach we initially introduce the dynamical polarization correction by calculating the self-energy part $\Sigma$ (2) in the second order of perturbation theory. With these corrections the DEM energy for the "3p↓" resonance state is $E^{DEM}_{3p\downarrow}$= - 0.49678 *Ry*. The corrected DEM $\widetilde{3p}\downarrow$ photoelectron wavefunction, which is actually a superposition of one-particle p↓ eigenfunctions of the SPHF Hamiltonian, is used for the further calculations. When the dynamic relaxation, the screened interaction of a photoelectron with the hole 2s↓, is taken into account by solving (6) and (7) with the corresponding screening diagrams (9), the oscillator strength of the 2s↓→ $\widetilde{3p}\downarrow$ phototransition turns out to be practically fully transferred into the εp↓ continuum. The oscillator strength $f^{DEM}_{2s\downarrow\to 3\tilde{p}\downarrow}$ is equal 0.0046 / 0.0049 (**r**- /∇-forms) when the dynamical relaxation and polarization corrections are included, *i.e.* one order of magnitude less than that in the one-particle approximation (and even two orders less without account of a polarization effect, due to the relaxation only). The calculations predict a strong resonance peak at the energy $E_{res}$=11.82 *Ry*, $\varepsilon_{res}$= 0.04 *Ry* with resonance width of Γ=0.02 *Ry*. It should be noted here that in what follows we use the SPHF value for threshold energies, so for a comparison with possible experiments the resonance region in fig.5 should be shifted, since our DEM correction to the threshold 2s↓ energy gives the more precise threshold value $E^{DEM}_{2s\downarrow}$=-11.408 *Ry*. The estimation for this value according to Koopman's theorem $E(\mathrm{Si}^{*})^{SPHF}_{tot} - E(\mathrm{Si}^{-})^{SPHF}_{tot}$ gives 11.112 *Ry*. Important also is the fact that in performing the present first calculations we keep the SPHF approximation for all other one-electron states except $\widetilde{3p}\downarrow$, when in perspective the nearest subshells from which the excitations $j \to k$ are important in the series (9) should be also "unfrozen" within the RPAE&DEM approach. However, these corrections would most likely only result in more precise determination of quantitative characteristics for the predicted resonance.

The middle panel of fig.5 shows the photoelectron phaseshift $\delta_1(\varepsilon)$: the p↓-wave phaseshift calculated in the one-particle SPHF approximation and also the corrections to it due to the dynamical relaxation and polarization within the DEM. One can see that the real part of correction (red line) demonstrates the characteristic behavior for a classical shape resonance. In the same time the reducible self-energy part and, correspondingly, the phaseshift also have an imaginary part (green line) that corresponds to a real bound state – a Feschbach resonance. Note, that here we do not consider the contribution of the RPAE correlations to affecting the phase shift, but only consider the DEM corrections. The latter is responsible for forming the mixed "shape-Feschbach" resonance.

When the interchannel interaction is included within the RPAE&DEM, the shape resonance peak stays practically unchanged, however this interaction affects strongly the partial photodetachment from the nearest 2s↑ subshell (destructive interference) and to a the lesser extent the partial cross sections for 2p electrons photodetachment (fig.4). One can see from fig.5 that the Fano-profiles due to interchannel interaction give a clear "trace" of the original bound state "3p↓", the Feschbach resonance, localized before the 2s↓ threshold. Correspondingly, the interaction with the other channels affects the oscillator strength for the 2s↓→ $\widetilde{3p}$↓ phototransition – $f^{RPAE\&DEM}_{2s\downarrow\rightarrow 3\tilde{p}\downarrow}$ = 0.020 (**r**- and $\nabla$-forms). The dual nature of the "3p↓ shape-Feschbach" resonance is evident also from the complex behavior of characteristics that are sensitive to partial profiles, such as the parameters of angular anisotropy $\beta_{2p\uparrow}$ and $\beta_{2p\downarrow}$ given in the upper panel of fig.5. The additional peculiarities of the Fano-profile type appear as a result in the total photodetachment cross section (lower panel of fig.5). However, the total $Si^-$ photodetachment cross section in the energy region under investigation is dominated by the strong resonance peak of complex "shape-Feschbach" nature just after the 2s threshold. We have to conclude from the above analyses that the dynamical relaxation is the most pronounced effect for the considered photoprocesses in the $Si^-$ strongly correlated many-electron system, however the total response of the whole system is shown to be emphatically collective.

**6. Summary**

In present work, photodetachment from the inner shells (2p,2s) of the open-shell negative ion $Si^-$ negative ion is studied theoretically and the emphasis is made for the importance of electron correlation effects for near 2s threshold photodetachment spectra. Due to the possibility of forming a quasi-bound "3p↓" state which leads to a resonance, the calculation for this region is of special interest. The initial estimations are made within the "frozen-field" RPAE and static relaxation (GRPAE) approximation and the calculated total cross sections have demonstrated typical interference structures, however without any prominent resonances. At the same time the existence of both limits of the "3p↓" state - as a bound state, or Feschbach resonance, in the "frozen-core" approximation and a quasi-bound state, shape resonance, in the static relaxation approximation - has been demonstrated. The relaxation of an atomic core results in the transformation of Feschbach resonance into shape one. Within the static approximation the effect of relaxation is overestimated. The dynamical description of the process (within the DEM) leads to substantial changing the resonance parameters. When the dynamical relaxation and polarization (DEM corrections) have been included, the resonance turns out to be of the complex mixed "shape-Feschbach" nature and manifests itself in a strong resonance peak just after the 2s threshold. Within the RPAE&DEM approach the calculations give the total photodetachment cross section with the sharp resonance structure near the 2s threshold that is analogous to the well-studied inner-shell resonance in $C^-$. However in the case of $Si^-$ due to interchannel interaction (RPAE correlations) the spectra is more complicated and reveals also additional Fano profiles in the energy region under investigation. The response of the strongly correlated system $Si^-$ to an electromagnetic field includes the electron correlations of all types considered within RPAE&DEM, however dynamical relaxation is shown to be the decisive factor in forming the complex strong "shape-Feschbach" resonance. The possible experimental evidence for the prominent resonance structure predicted in this work could be of a great interest.

**Acknowledgments**

The author thanks Prof. C. W. Walter and Prof. V. K. Ivanov for valuable discussion on the subject and the Department of Physics and Astronomy at the Aarhus University where the part of computer-resources-consuming calculations have been performed for hospitality.

**Figure captions**

**Figure 1.** Partial cross sections for the inner 2s and 2p shells of $Si^-$ calculated in the “frozen field” approximation: black (**r**-form) and red (∇-form) dashed lines – in the one-particle SPHF approximation with the wavefunctions of photoelectron founded in the field of “frozen” core; blue (**r**-form) and green (∇-form) lines – with account of the RPAE interactions between six channels.

**Figure 2.** Partial cross sections for the inner 2s and 2p shells of $Si^-$ calculated in the static relaxation approximation: black (**r**-form) and red (∇-form) dashed lines – in the one-particle SPHF approximation with the wavefunctions of photoelectron founded in the field of completely rearranged core; green (**r**-form) and blue (∇-form) lines – with account of the GRPAE interactions between six channels. Inset: p↓-photoelectron phaseshift $\delta_{1\downarrow}^{GRPAE}(\varepsilon)$ corresponding to the one-particle approximation wavefunctions within the GRPAE.

**Figure 3.** Total photodetachment cross section in the “frozen field” and the static relaxation approximations. The partial cross sections from figs. 1 and 2 and their sums are also depicted.

**Figure 4.** Partial cross sections for the inner 2s and 2p shells of $Si^-$ calculated within the RPAE&DEM method: black (**r**-form) and red (∇-form) dashed lines – in the one-particle SPHF approximation with the wavefunctions of photoelectron founded in the field of “frozen” core; blue (**r**-form) and green (∇-form) lines – with account of the dynamical relaxation, polarization and RPAE corrections.

**Figure 5.** The results of calculations within the RPAE&DEM approach – the dynamical relaxation, polarization and interchannel interaction are included.
Upper panel: The photoelectron angular asymmetry parameter for photodetachment from 2p shells of $Si^-$ in **r**-form (black) and ∇-form (red) of a dipole operator.
Middle panel: The scattering phaseshift for εp↓-photoelectron. The phaseshift $\delta_1(\varepsilon)$ is divided into three components: the HF phaseshift $\delta_1^{HF}(\varepsilon)$ (black line) and an additional phaseshift $\Delta\delta_1(\varepsilon)$ due to the dynamical relaxation and polarization corrections accounted within the within the DEM: $\mathcal{R}e\Delta\delta_1$ (red line) and $\Im m\Delta\delta_1$ (green line); dotted lines correspond to consideration only a critical for a formation of the resonance dynamical relaxation effect without account for polarization.
Lower panel: The total photodetachment cross section in the range of the inner 2s and 2p shells threshold energies for $Si^-$ calculated within the RPAE&DEM method. The partial cross sections from fig. 4 are also depicted.

# Si⁻. “FROZEN-FIELD” (RPAE). PARTIAL CROSS SECTIONS.

**2s↓→εp↓**

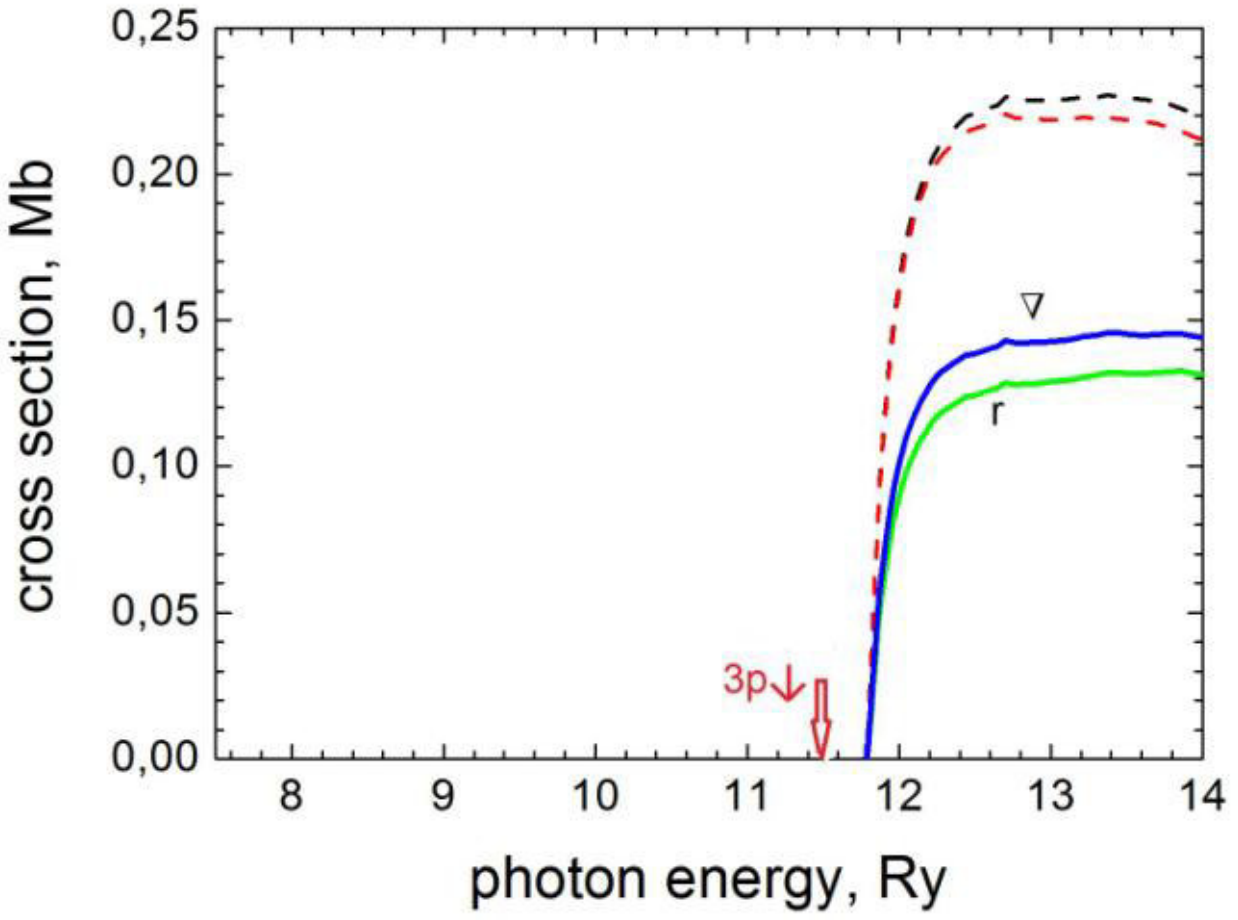

**2s↑→εp↑**

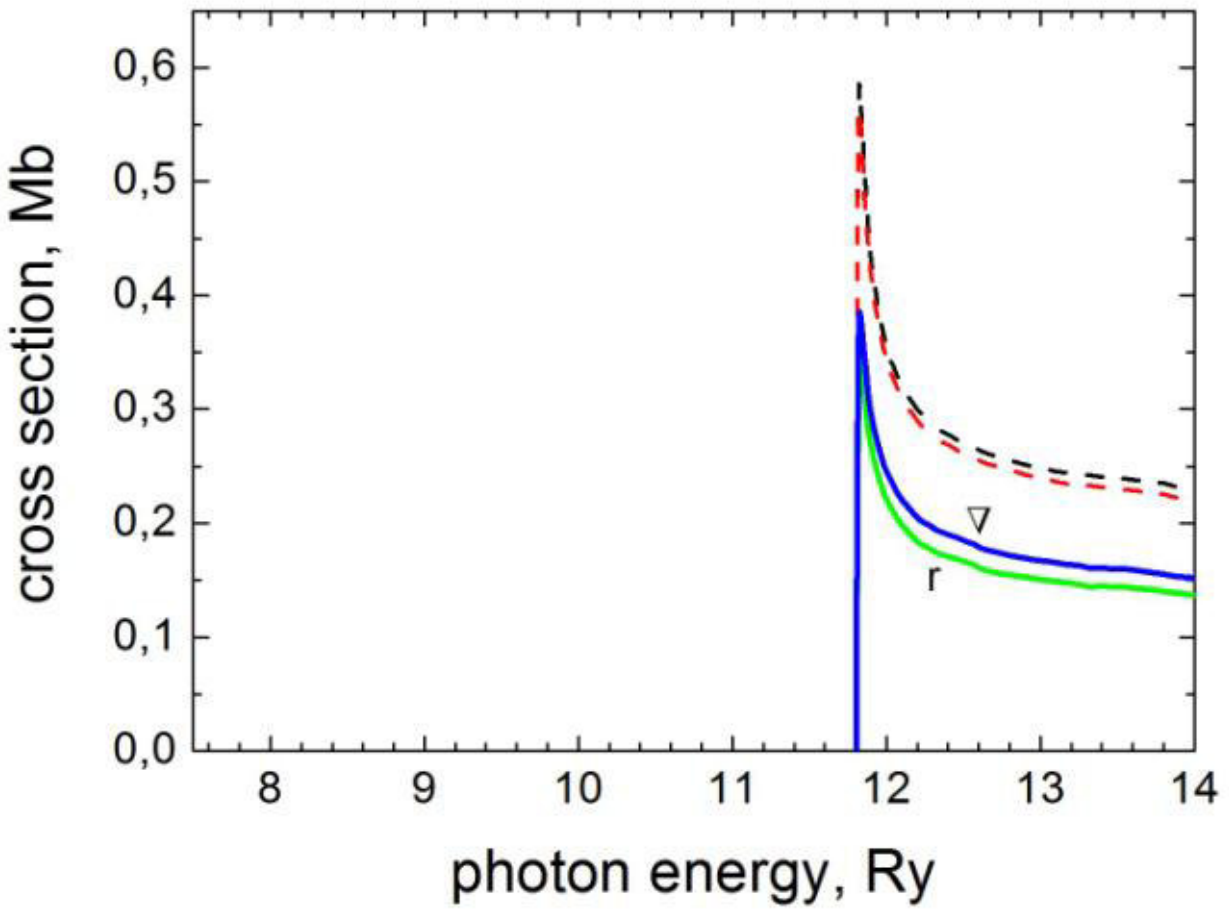

**2p↑→εd↑**

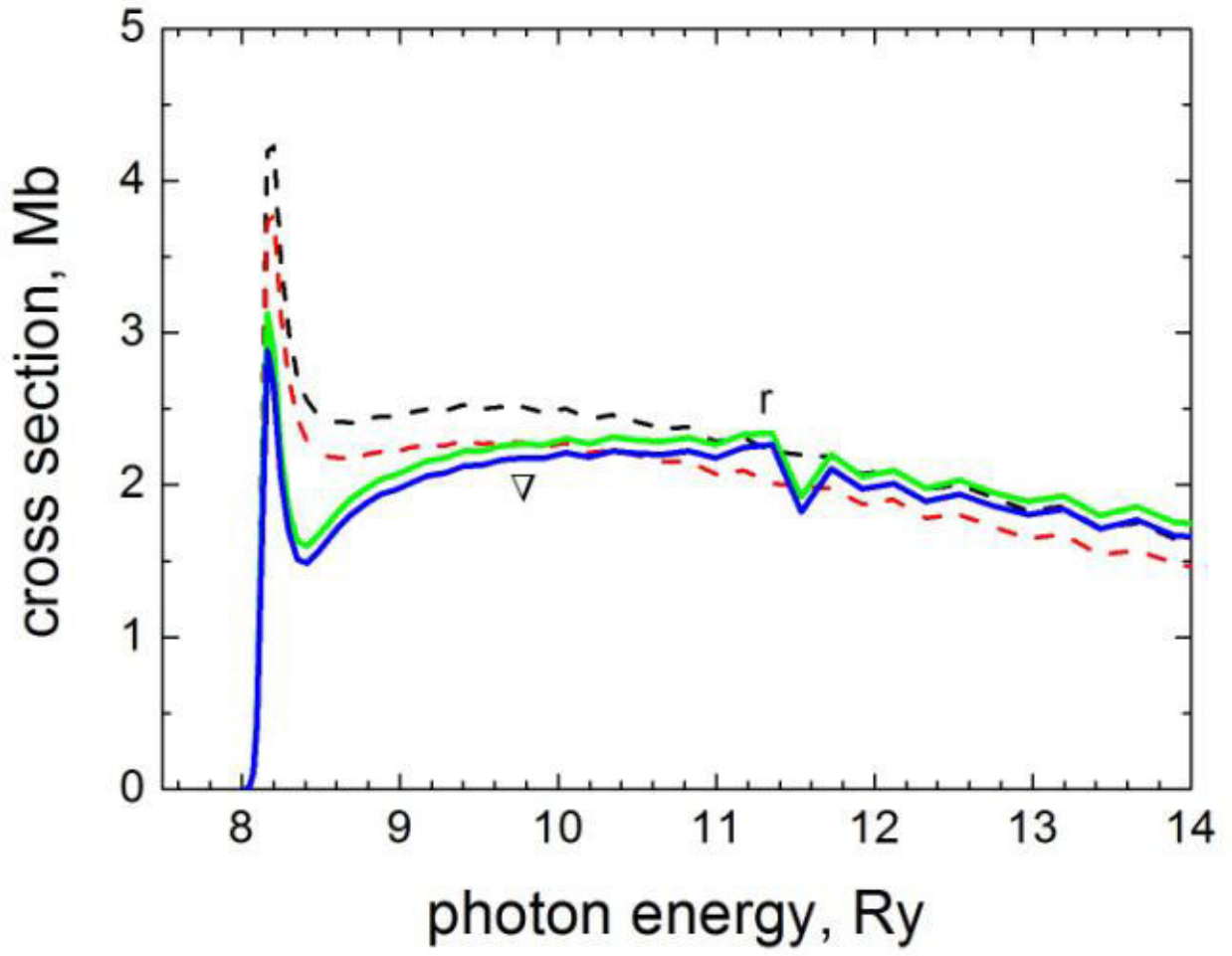

**2p↑→εs↑**

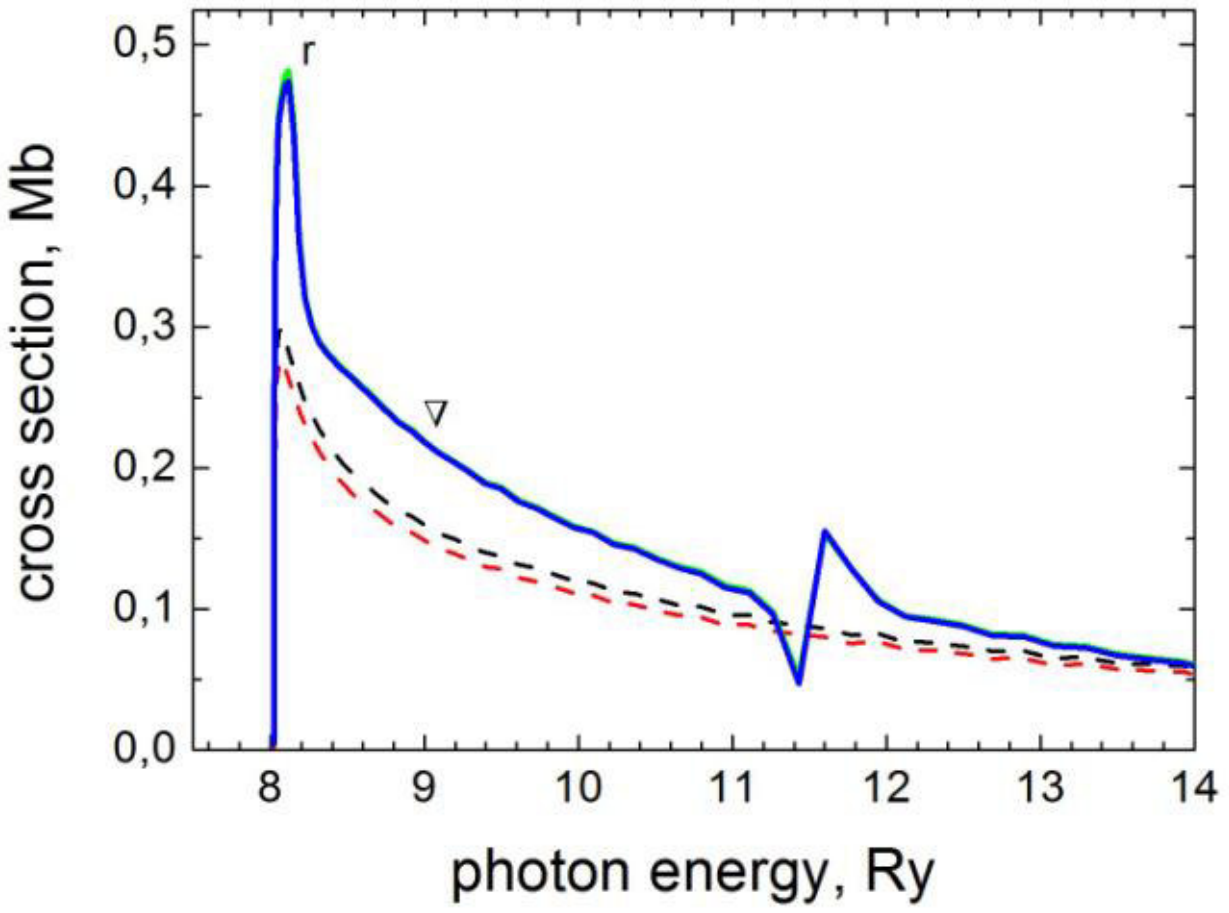

**2p↓→εd↓**

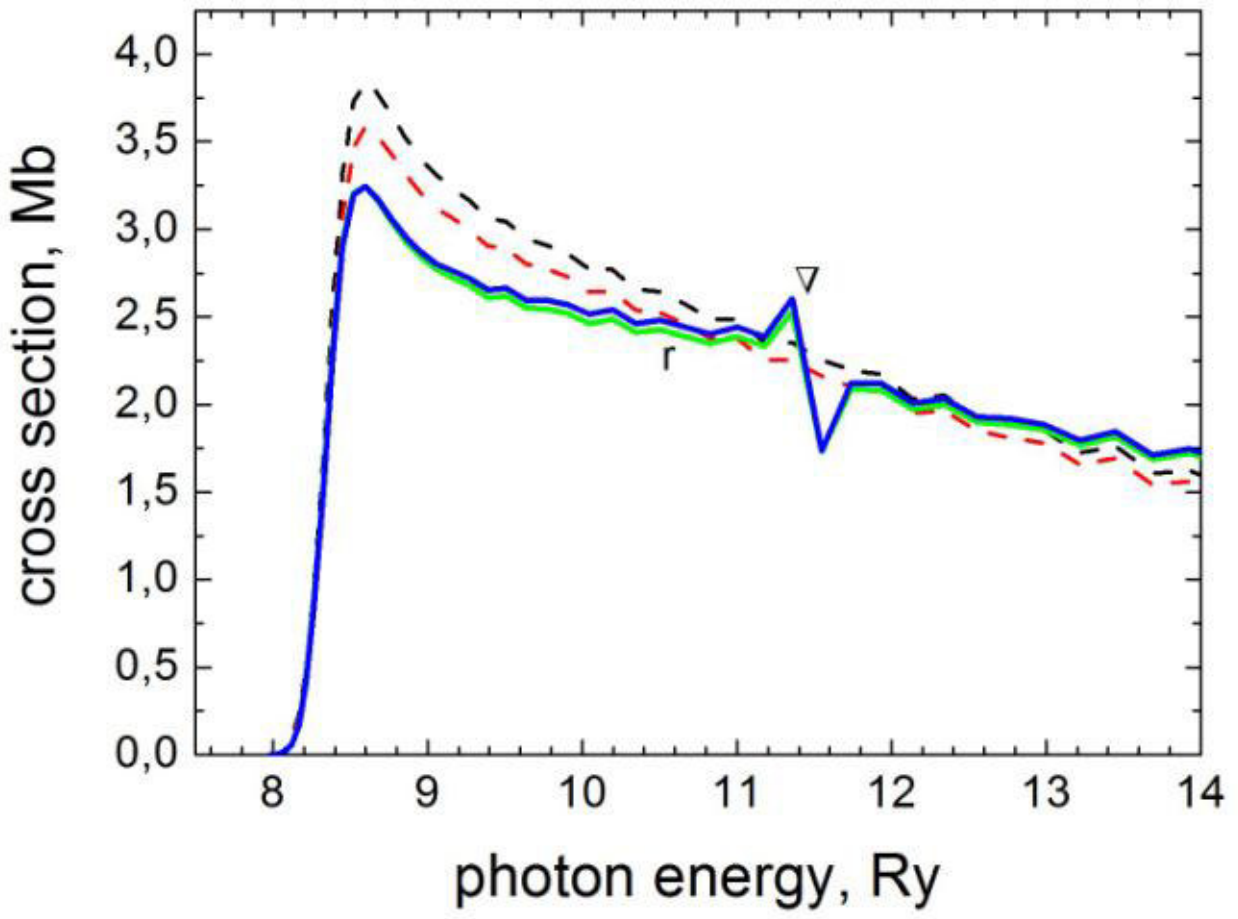

**2p↓→εs↓**

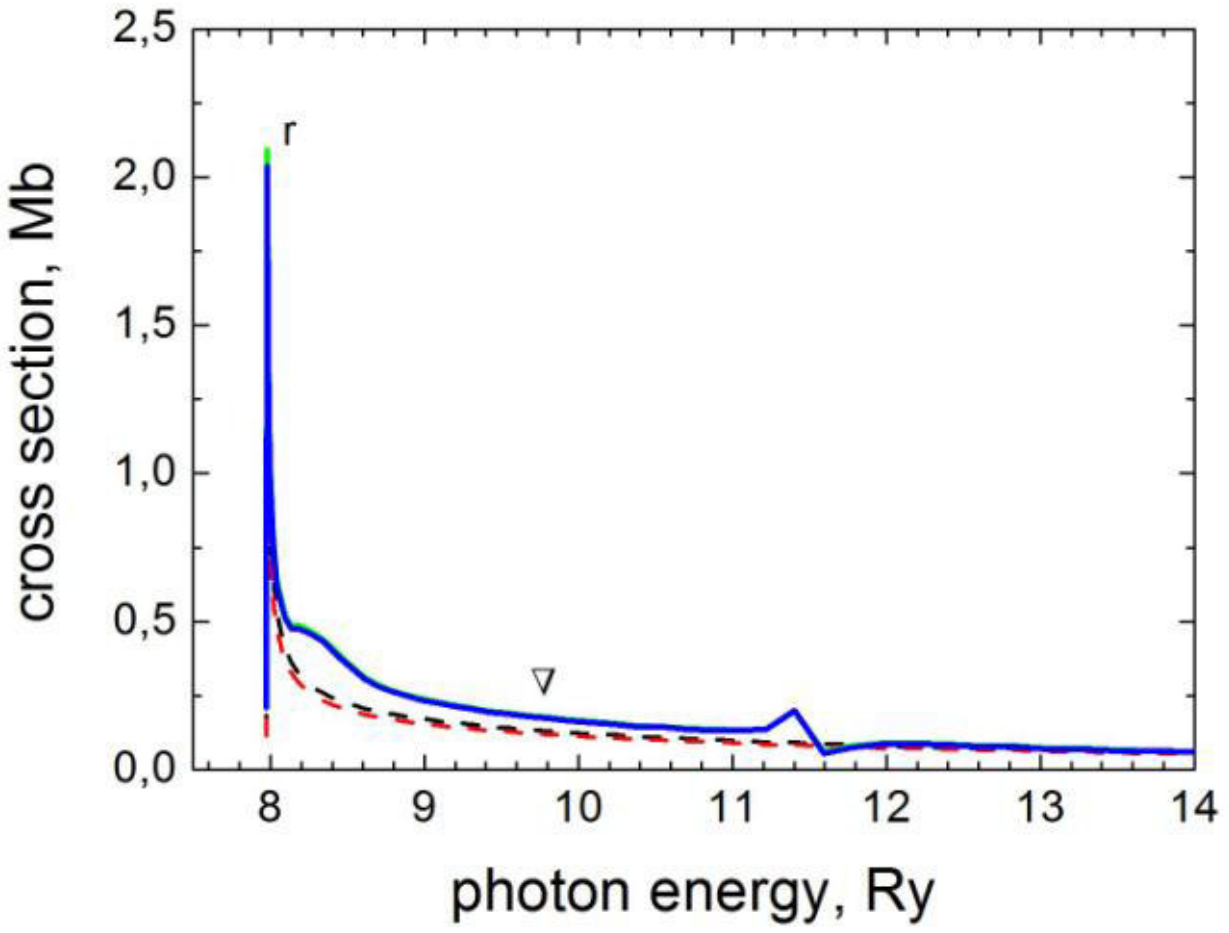

# Si⁻. STATIC RELAXATION (GRPAE). PARTIAL CROSS SECTIONS.

**2s↓→εp↓**

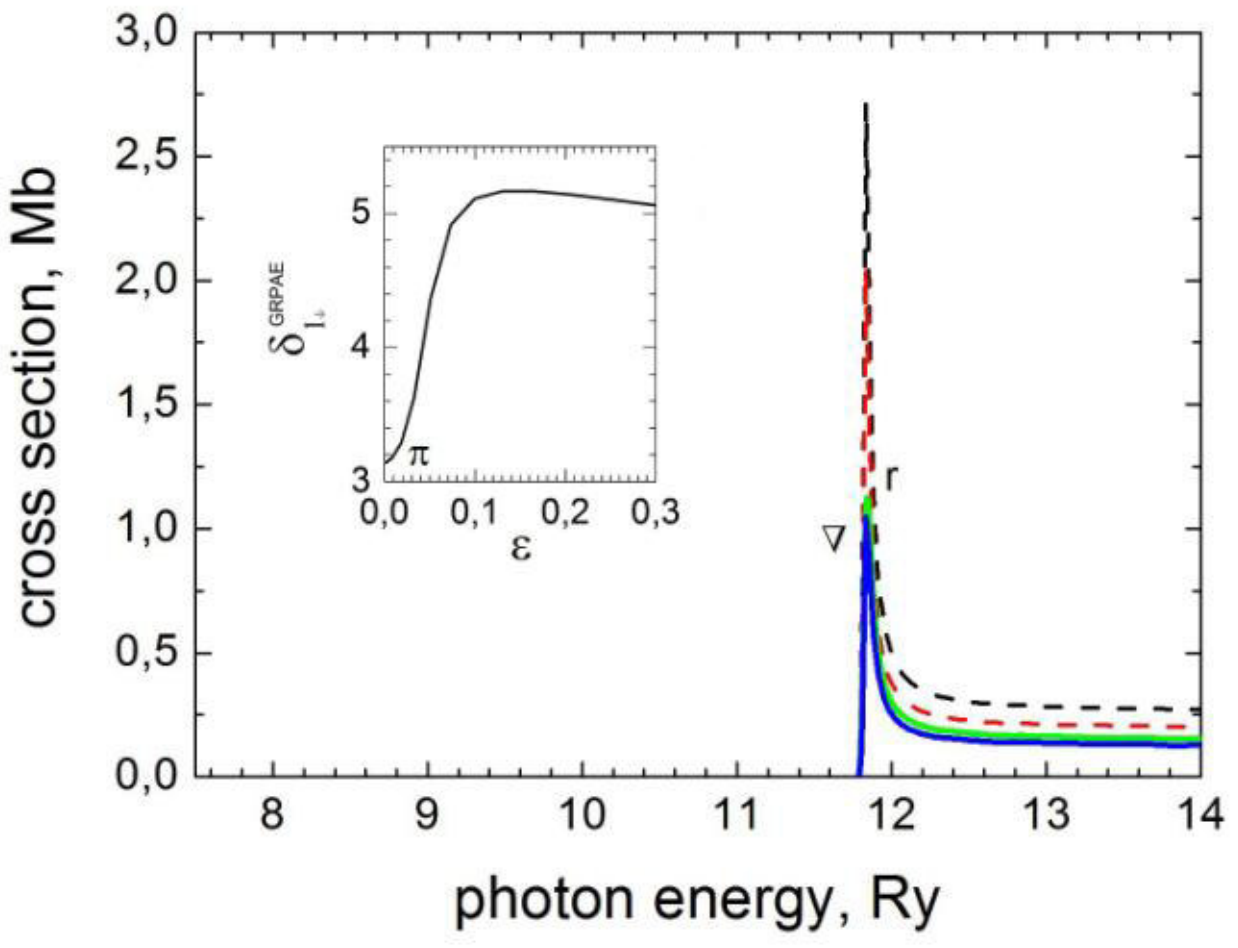

**2s↑→εp↑**

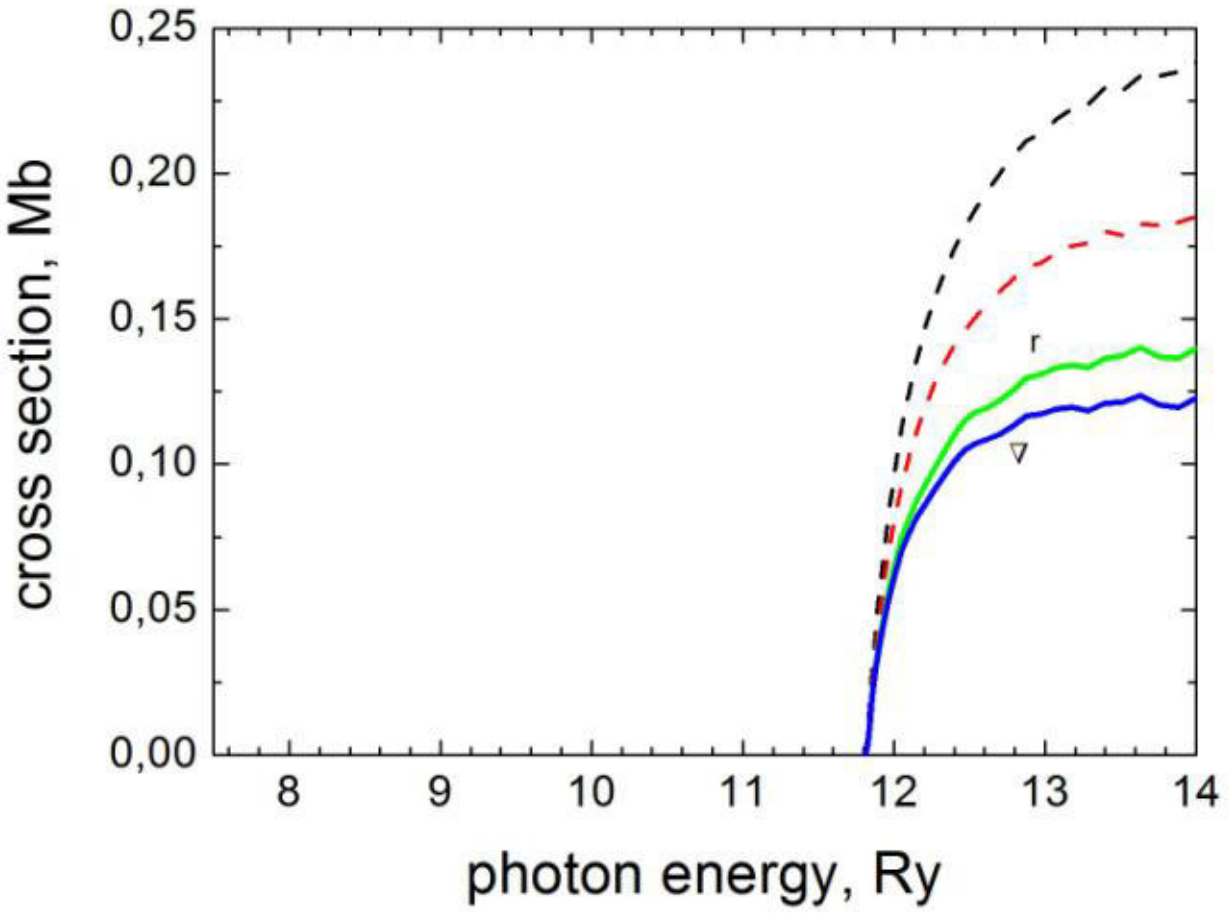

**2p↑→εd↑**

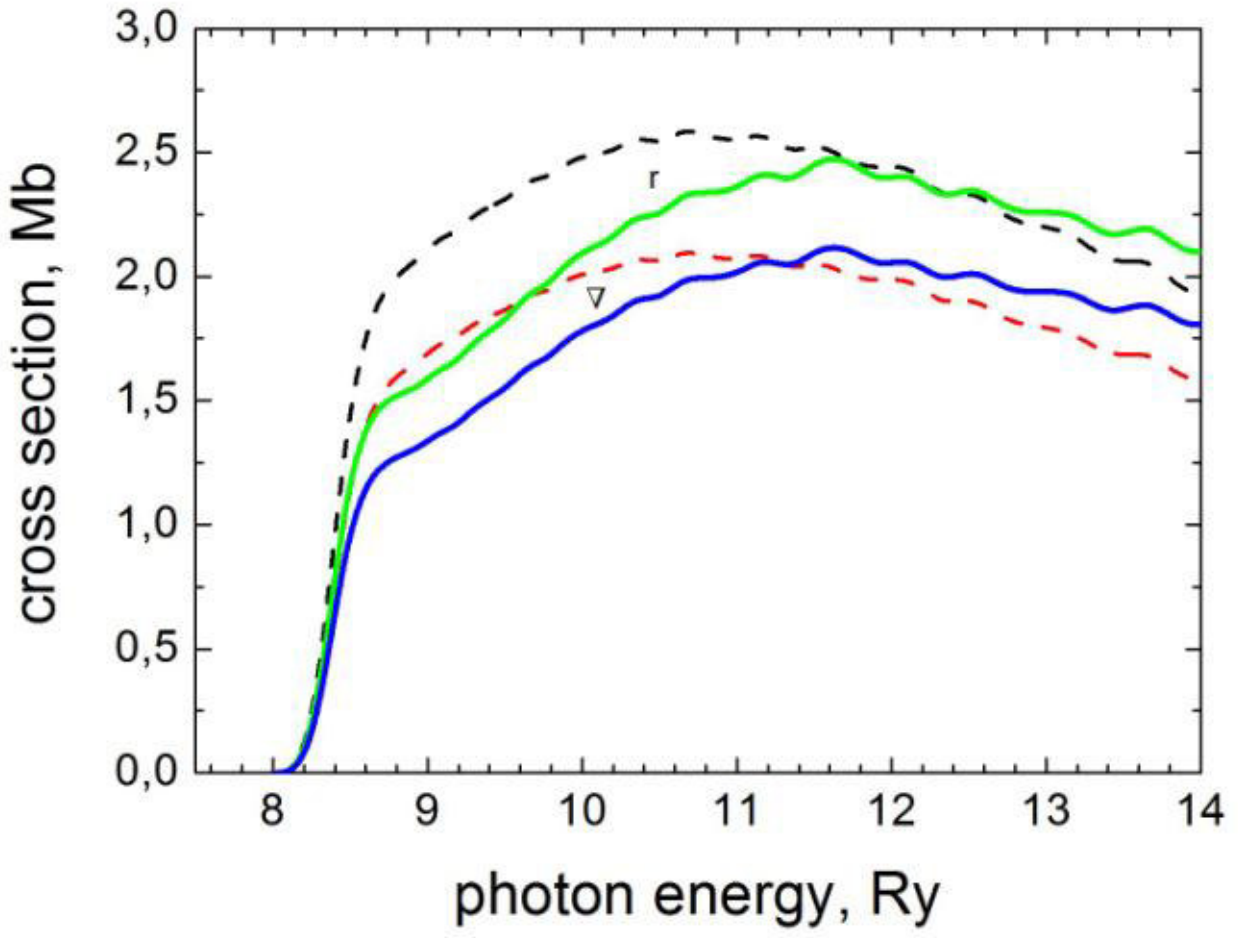

**2p↑→εs↑**

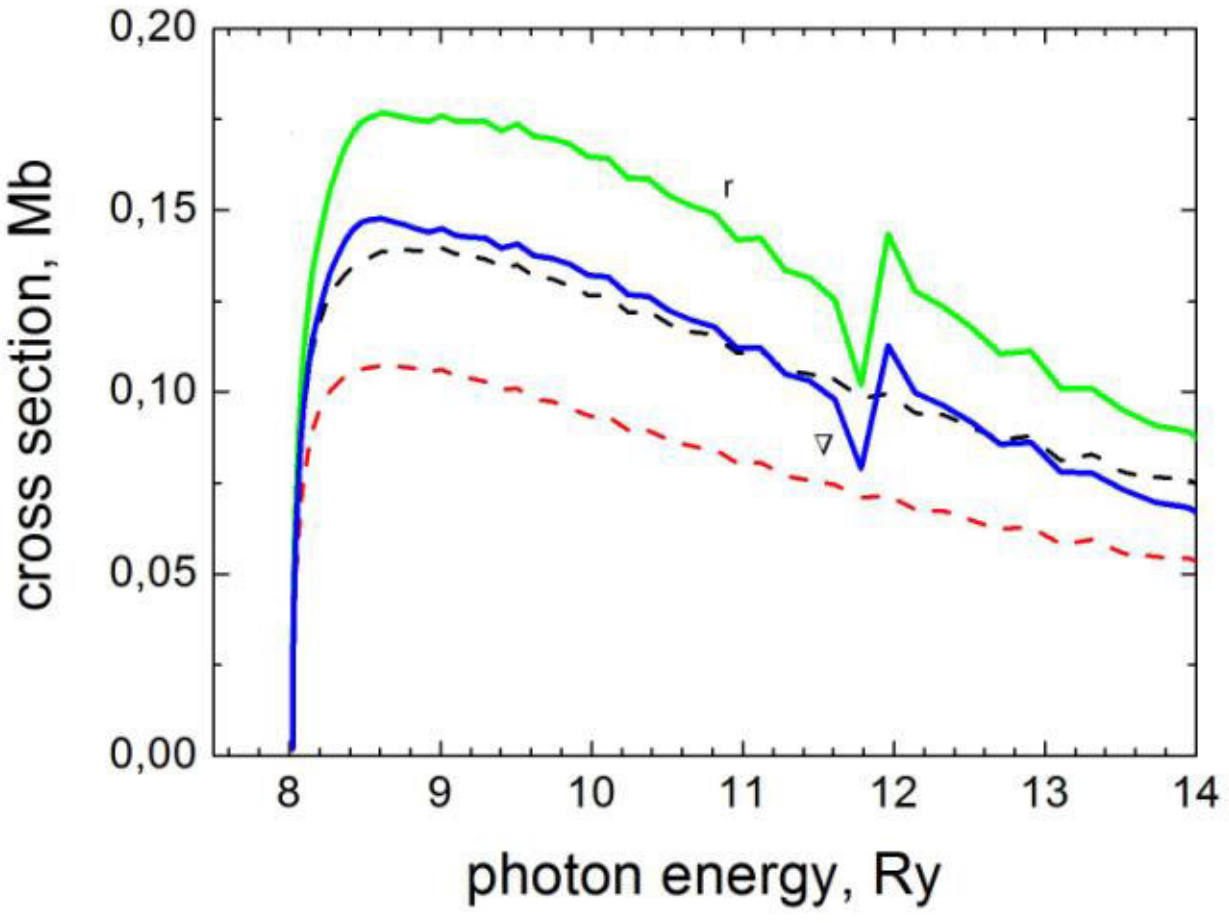

**2p↓→εd↓**

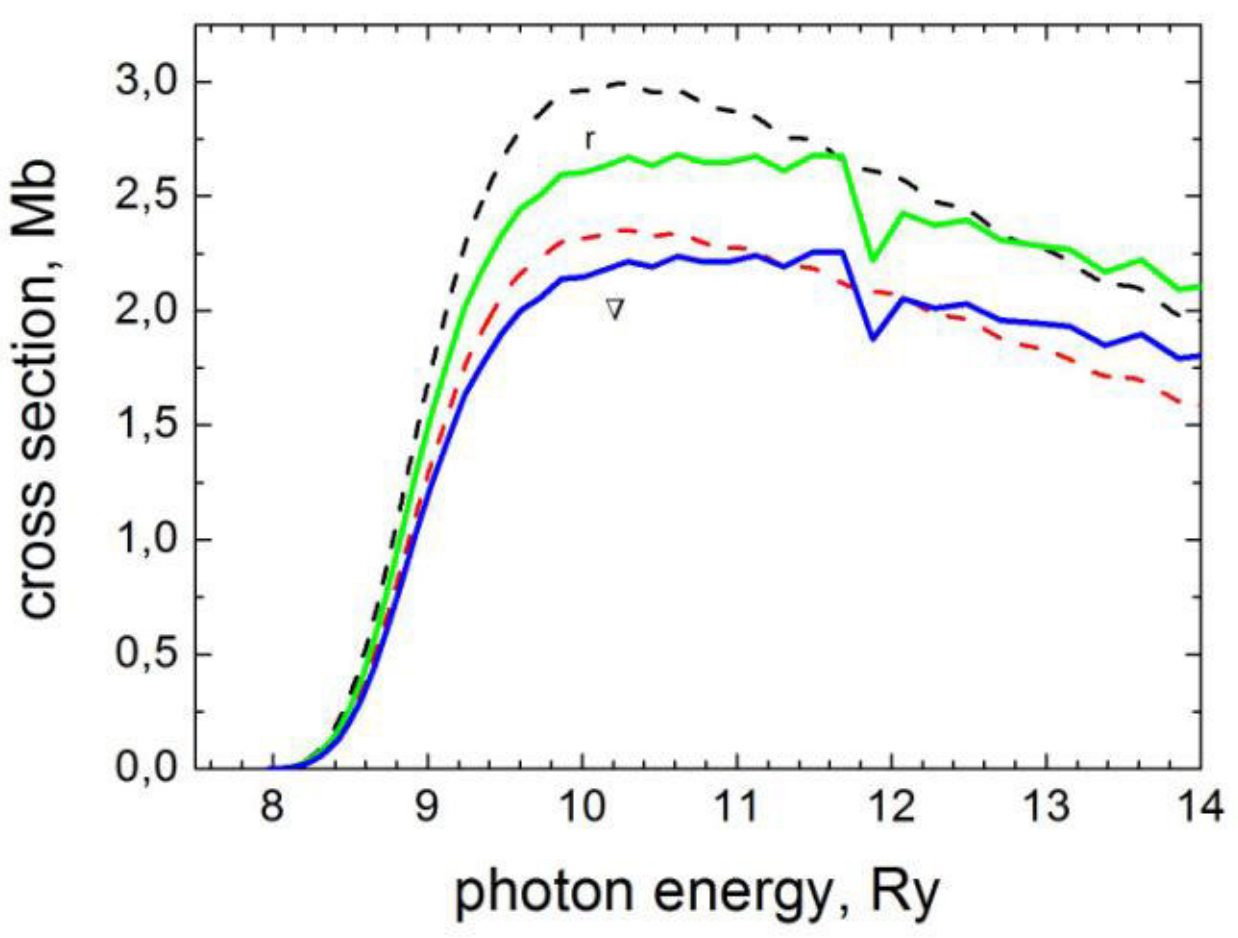

**2p↓→εs↓**

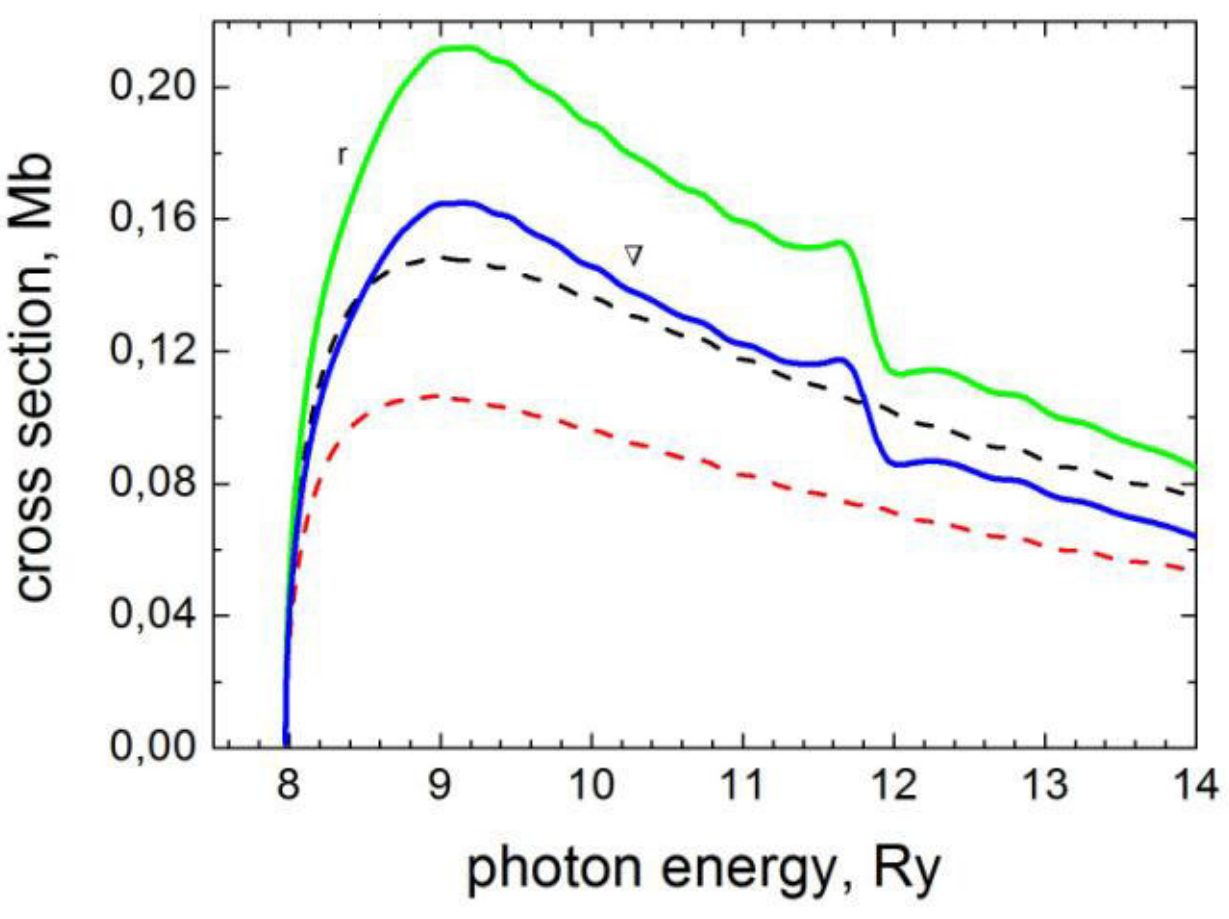

# Si⁻. STATIC & "FROZEN-FIELD" APPROXIMATIONS. GRPAE *vs* RPAE.

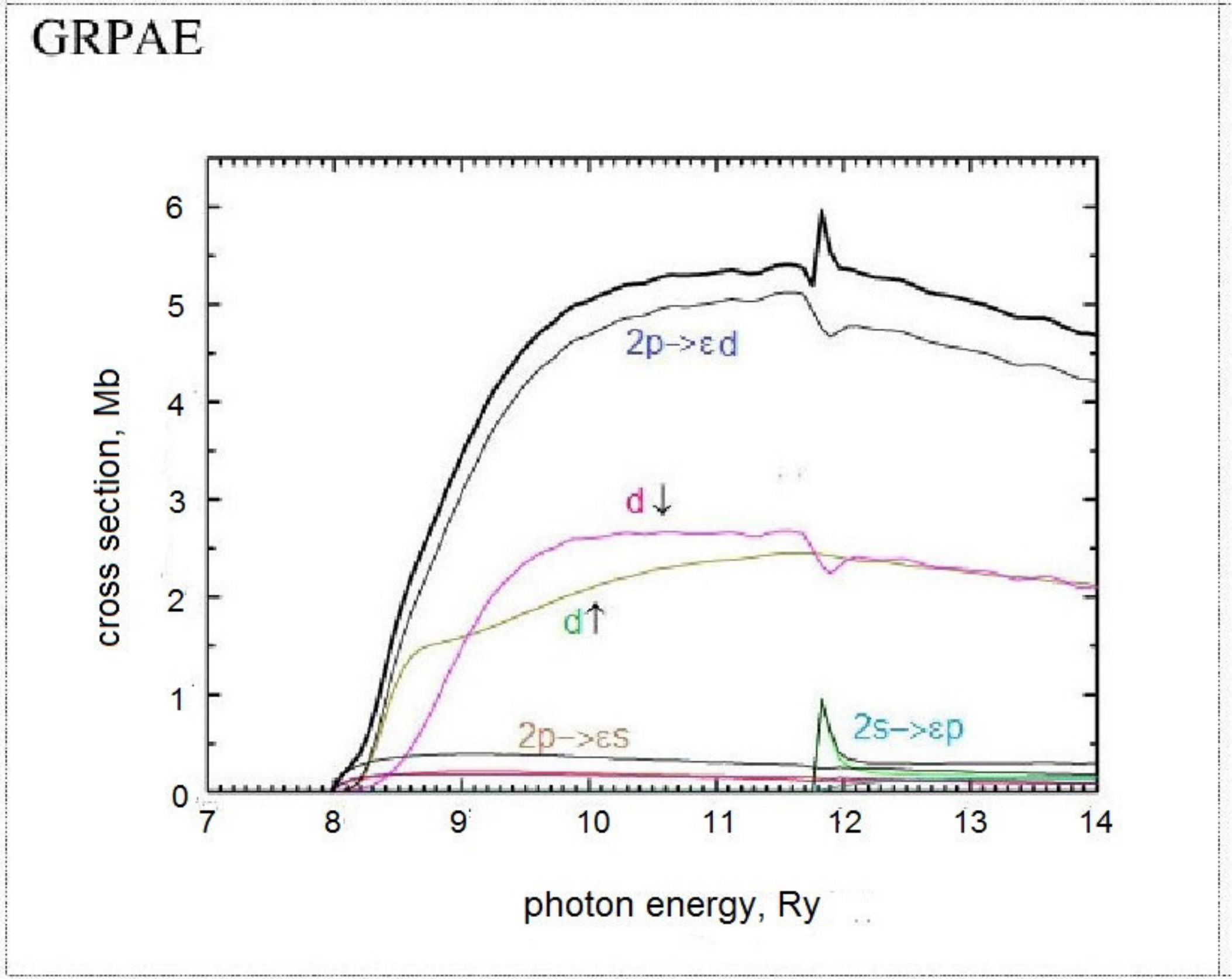

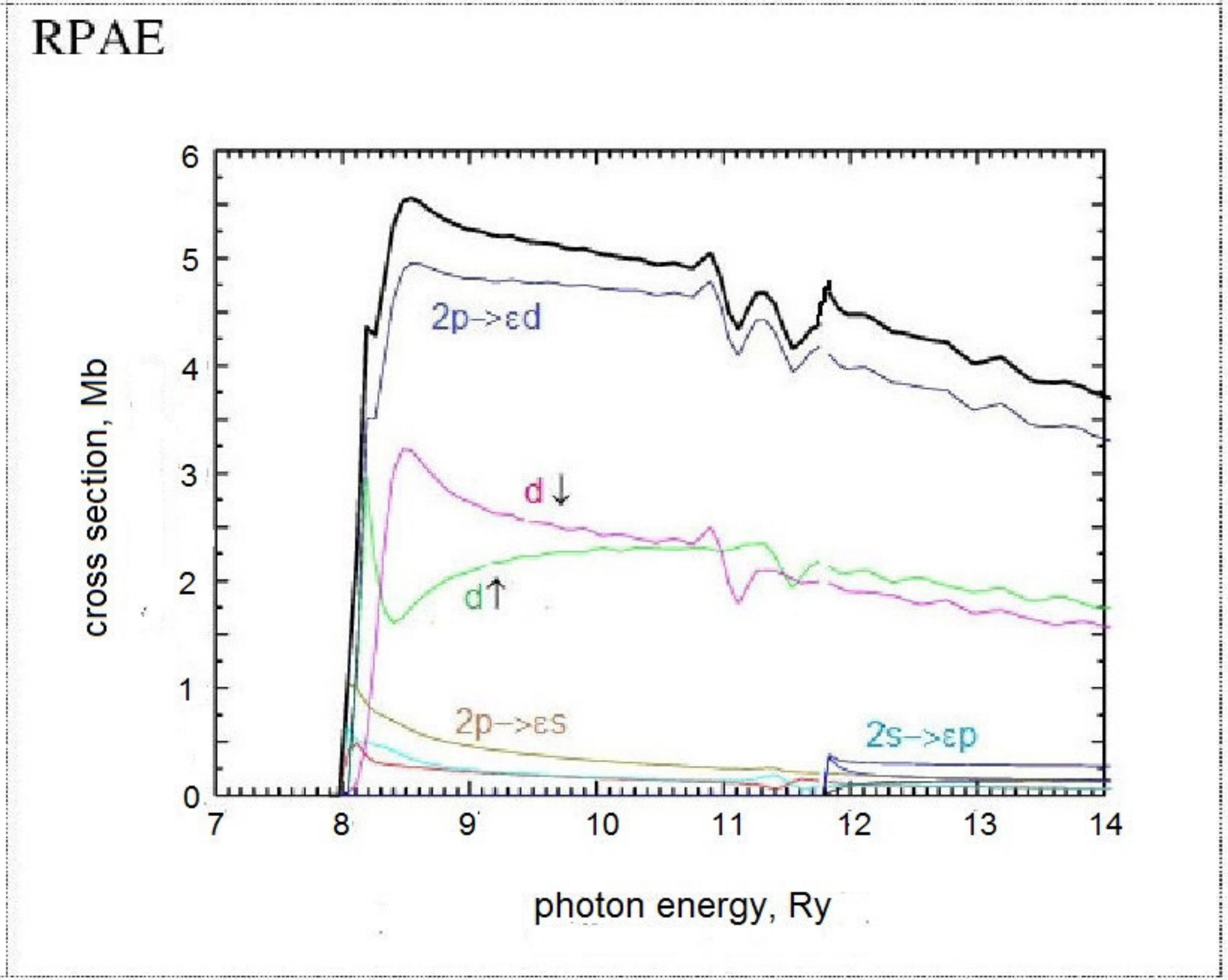

# Si⁻. RPAE&DEM. PARTIAL CROSS SECTIONS.

**2s↓→εp↓**

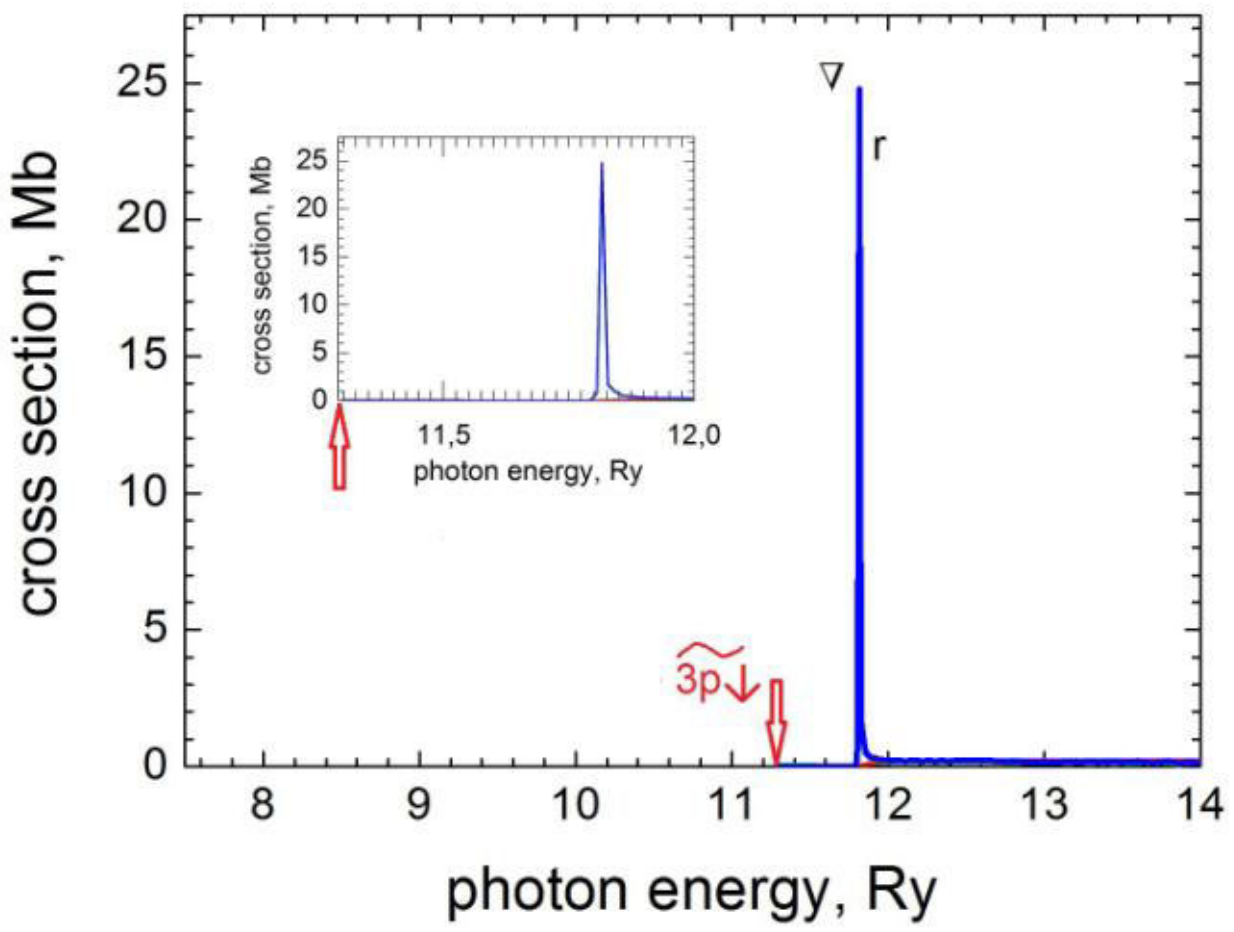

**2s↑→εp↑**

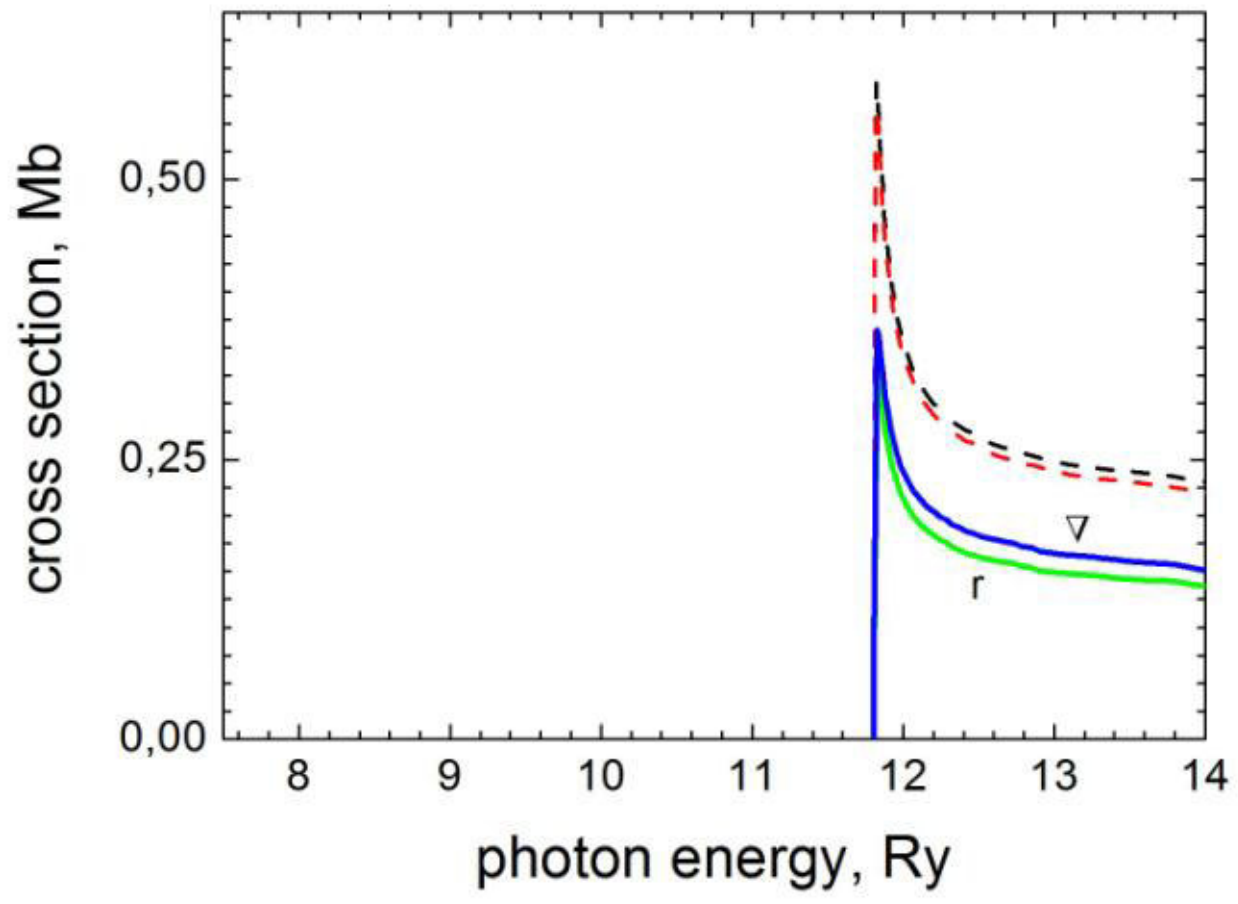

**2p↑→εd↑**

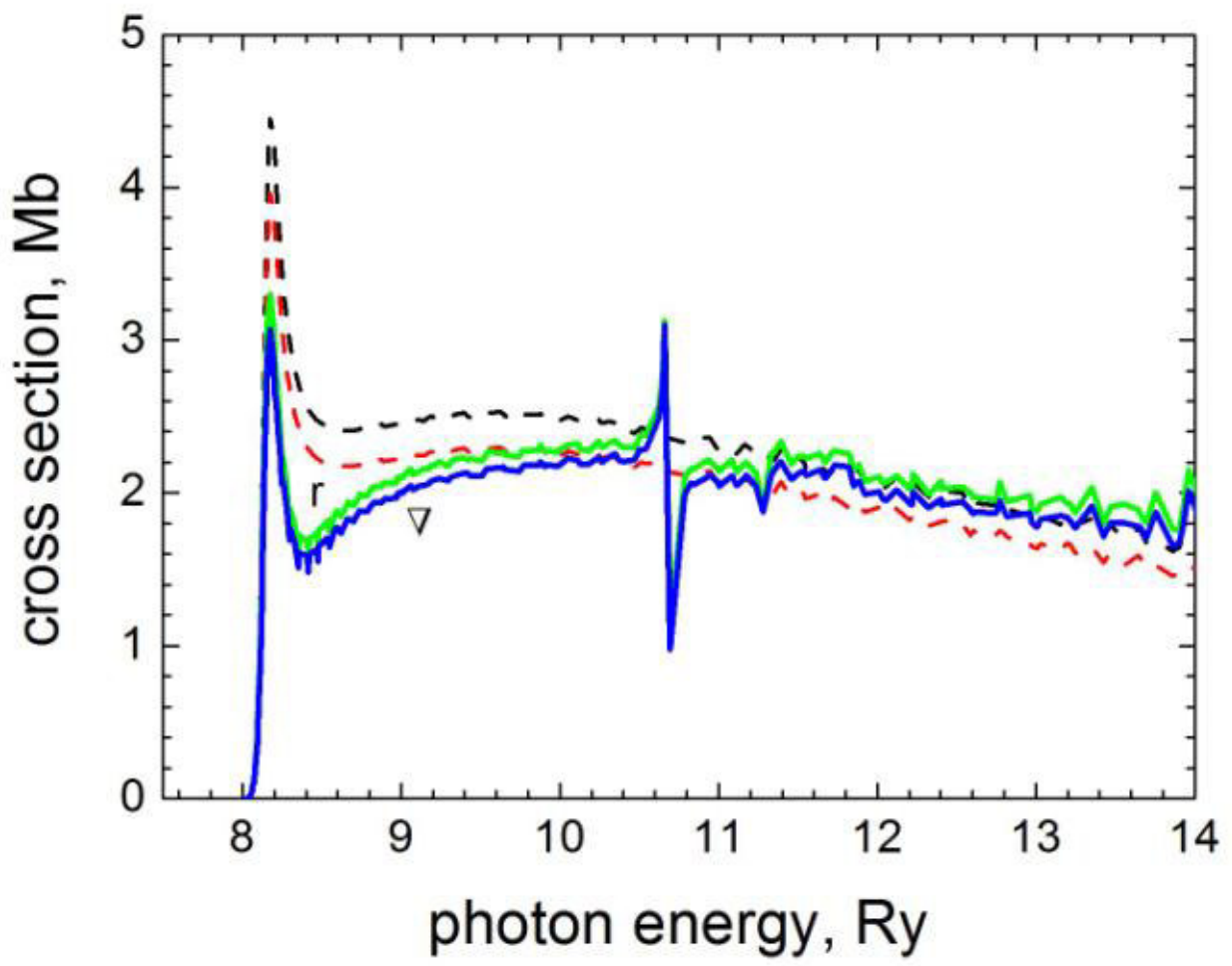

**2p↑→εs↑**

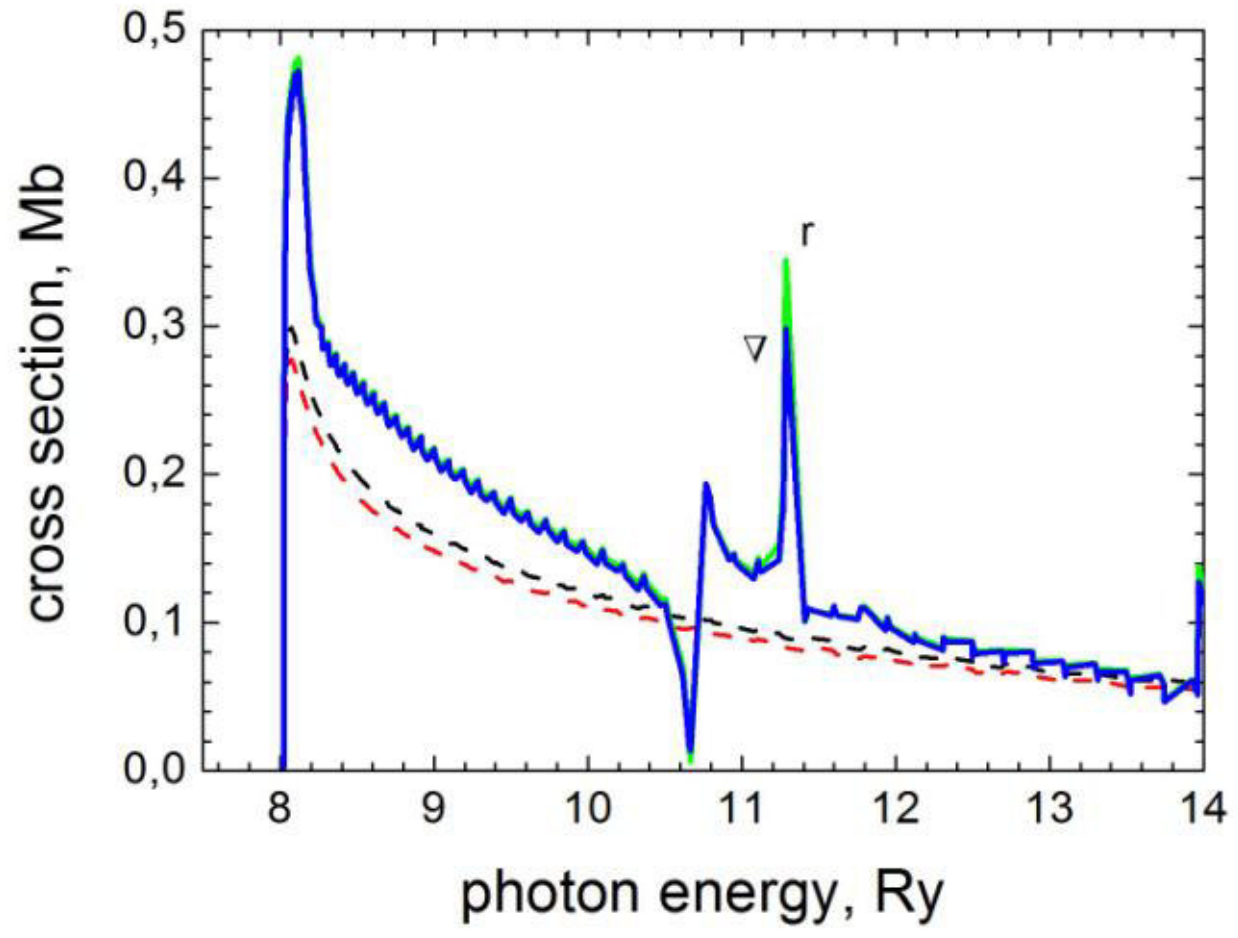

**2p↓→εd↓**

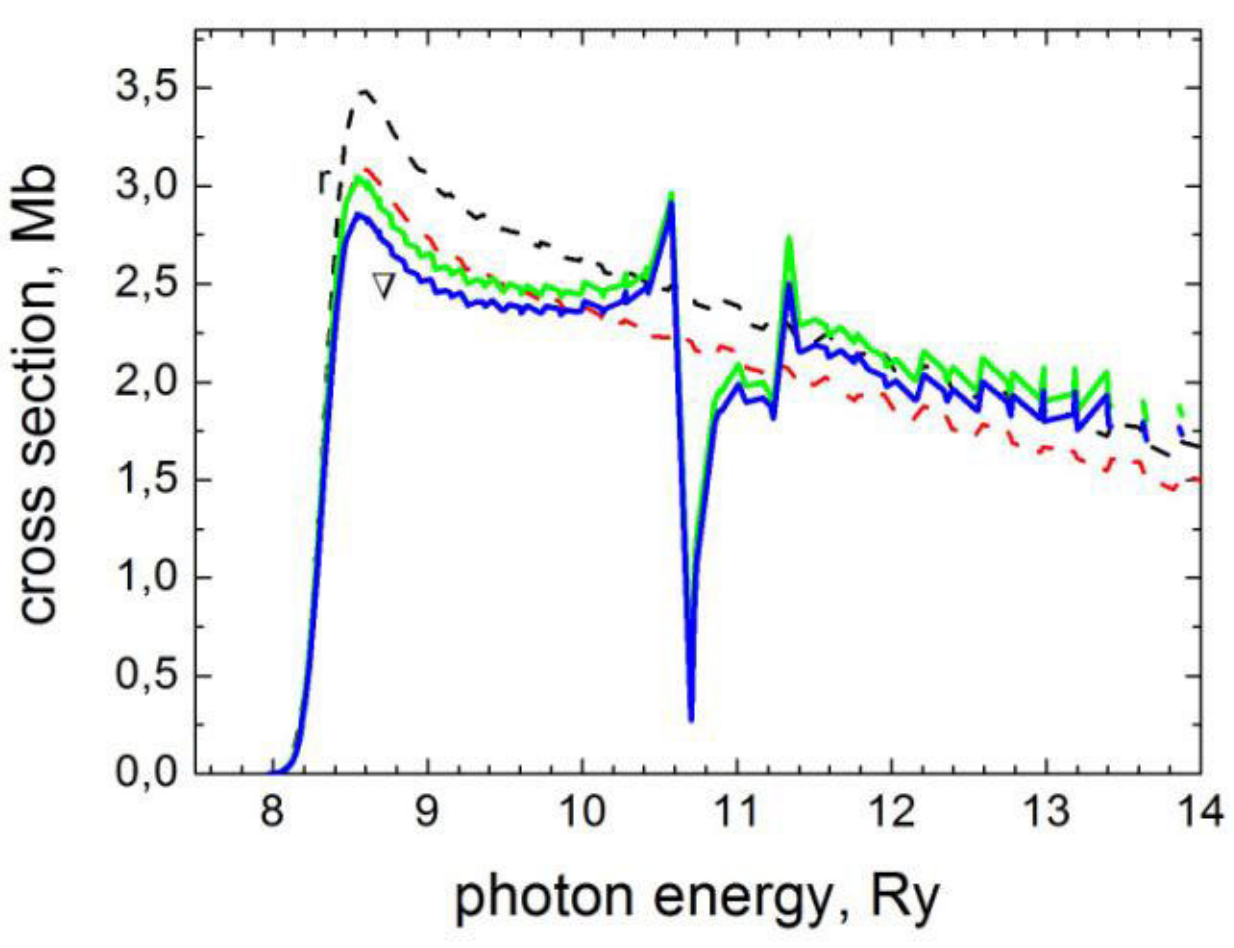

**2p↓→εs↓**

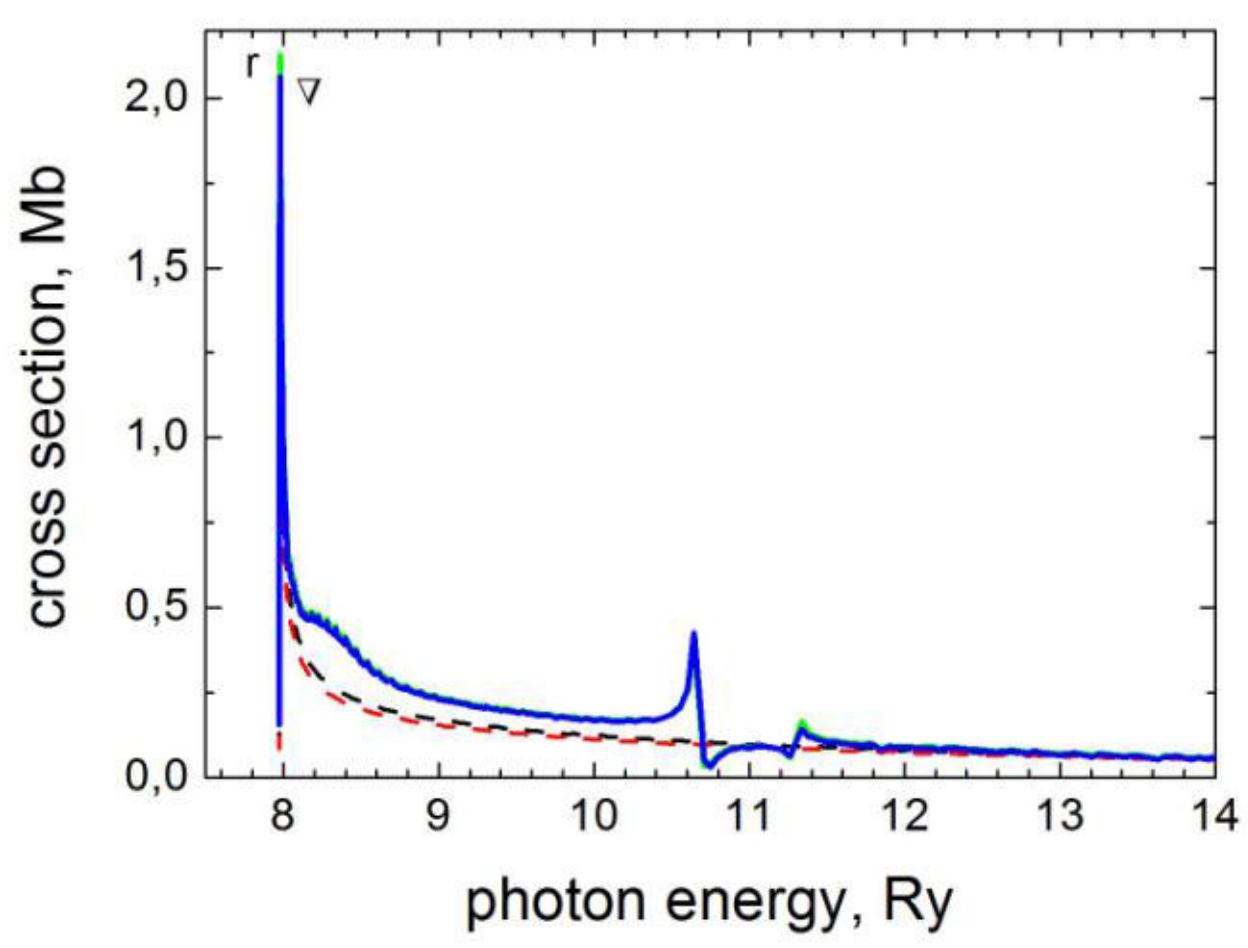

# $Si^-$. RPAE&DEM.

$\beta_{2p\downarrow}$

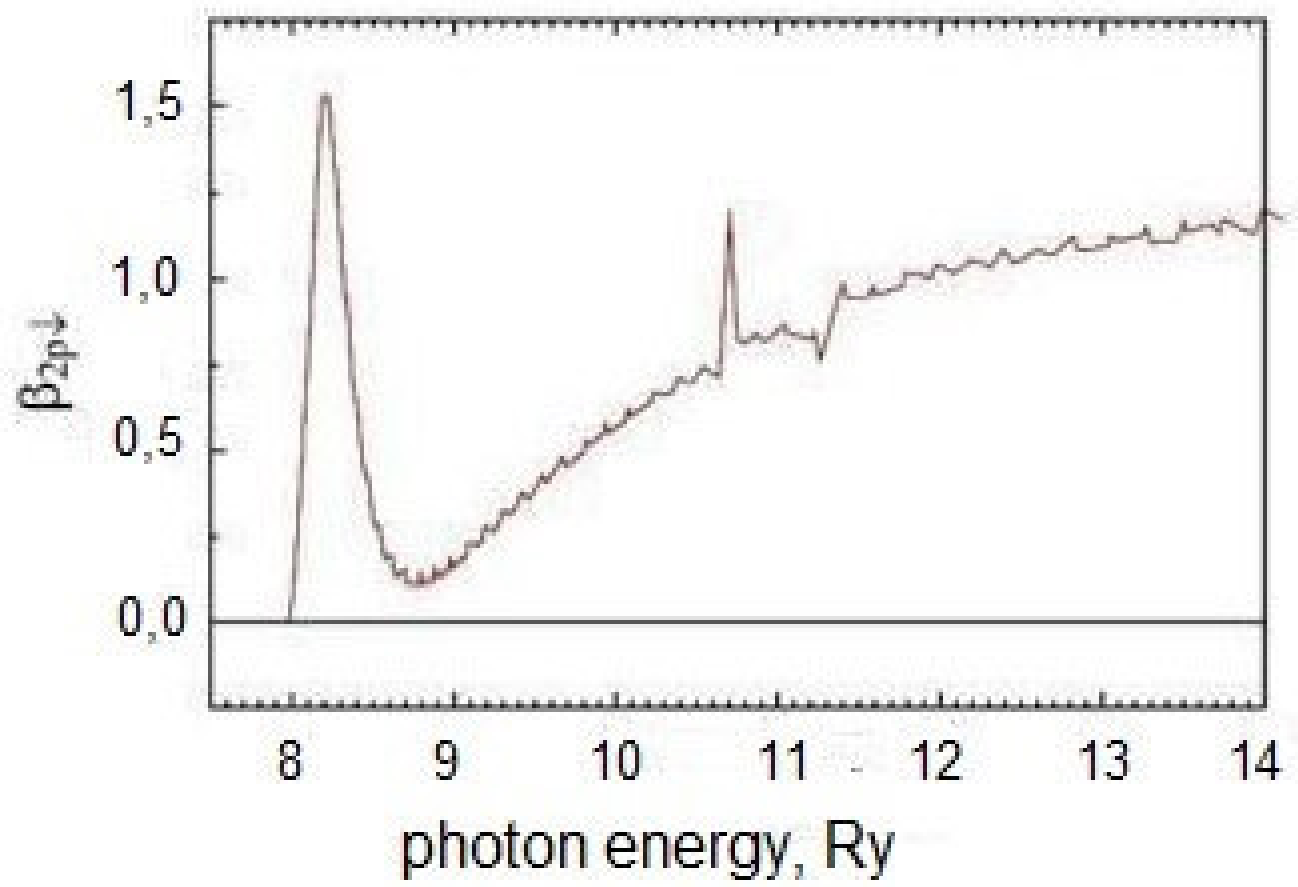

$\beta_{2p\uparrow}$

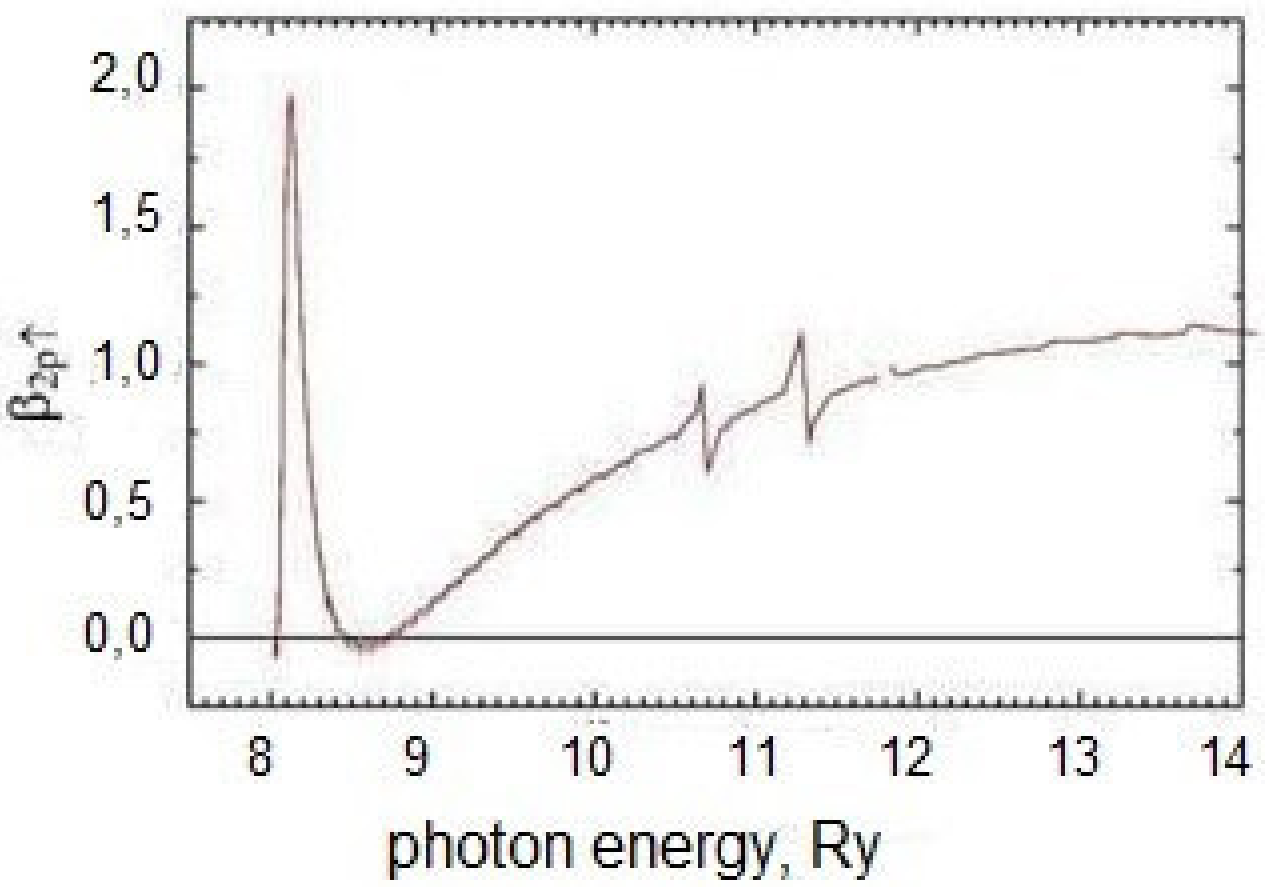

photoelectron phaseshift (p↓-wave)

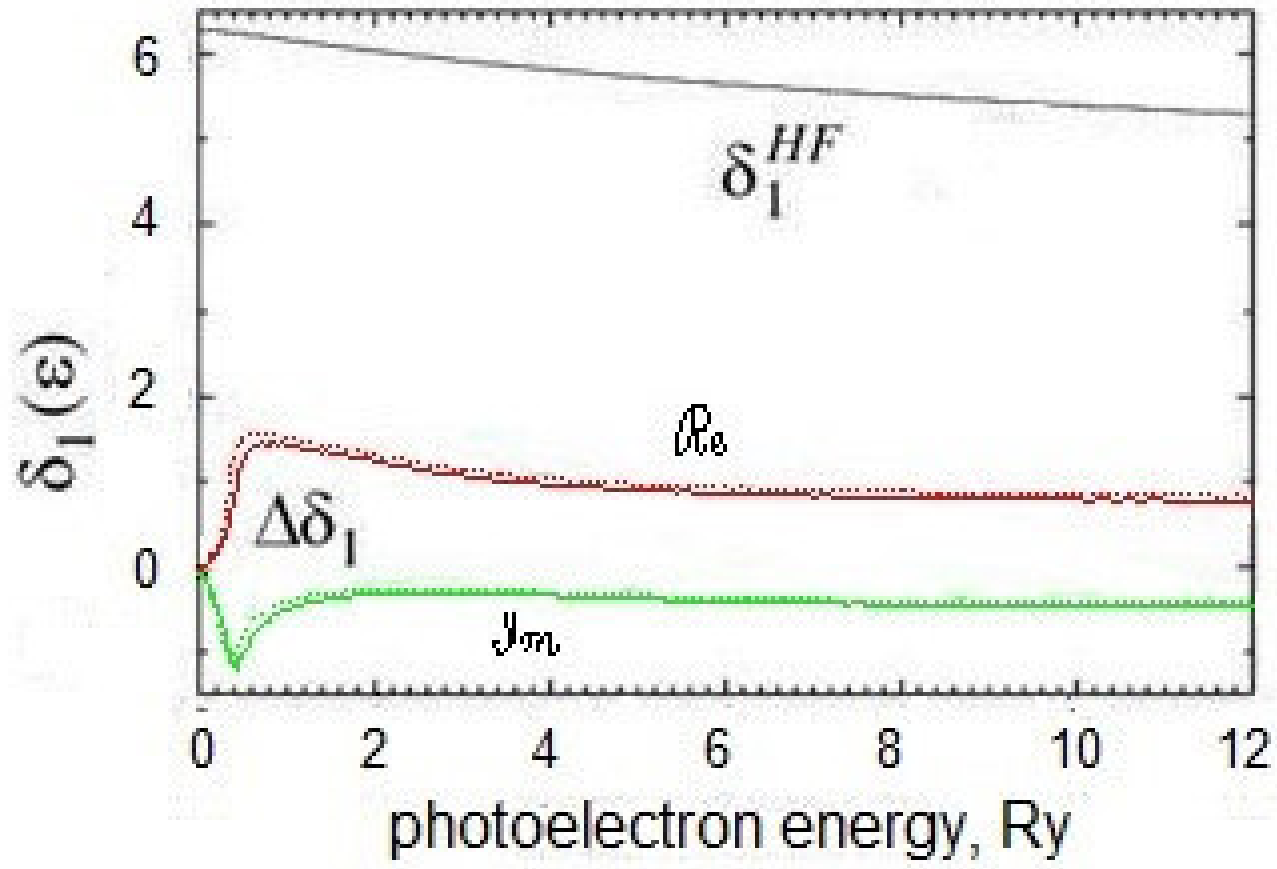

$\delta_1^{HF}$ - SPHF approximations

$\Delta\delta_1$ - corrections to it within the DEM:

$tan\,\Delta\delta_l(\varepsilon) = -\pi\langle\varepsilon|\tilde{\Sigma}_\varepsilon|\varepsilon\rangle$ (Amusia 1990)

$$\delta_1(\varepsilon) = \delta_1^{HF} + \Delta\delta_1(\varepsilon)$$

RPAE&DEM

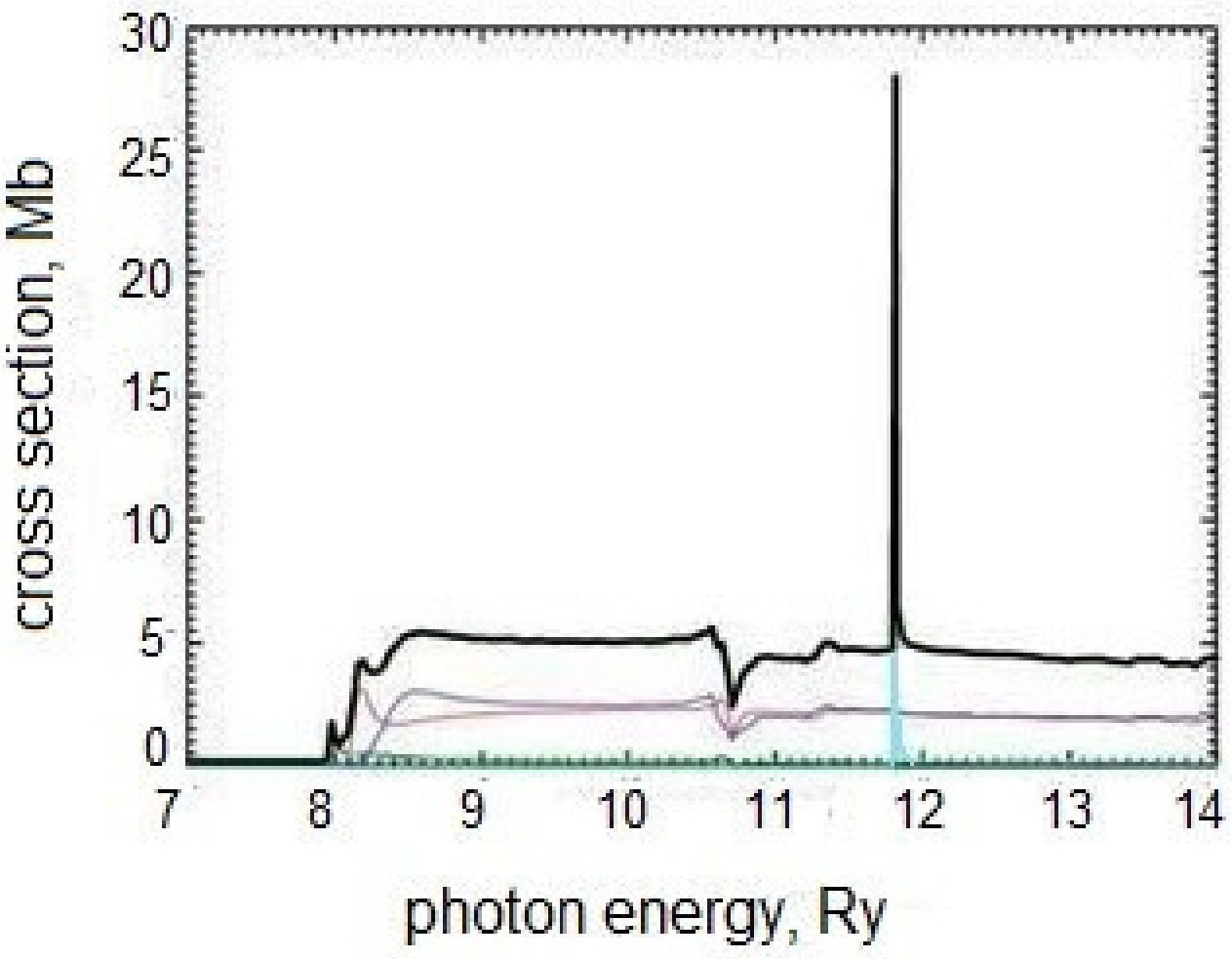

partial and total cross sections

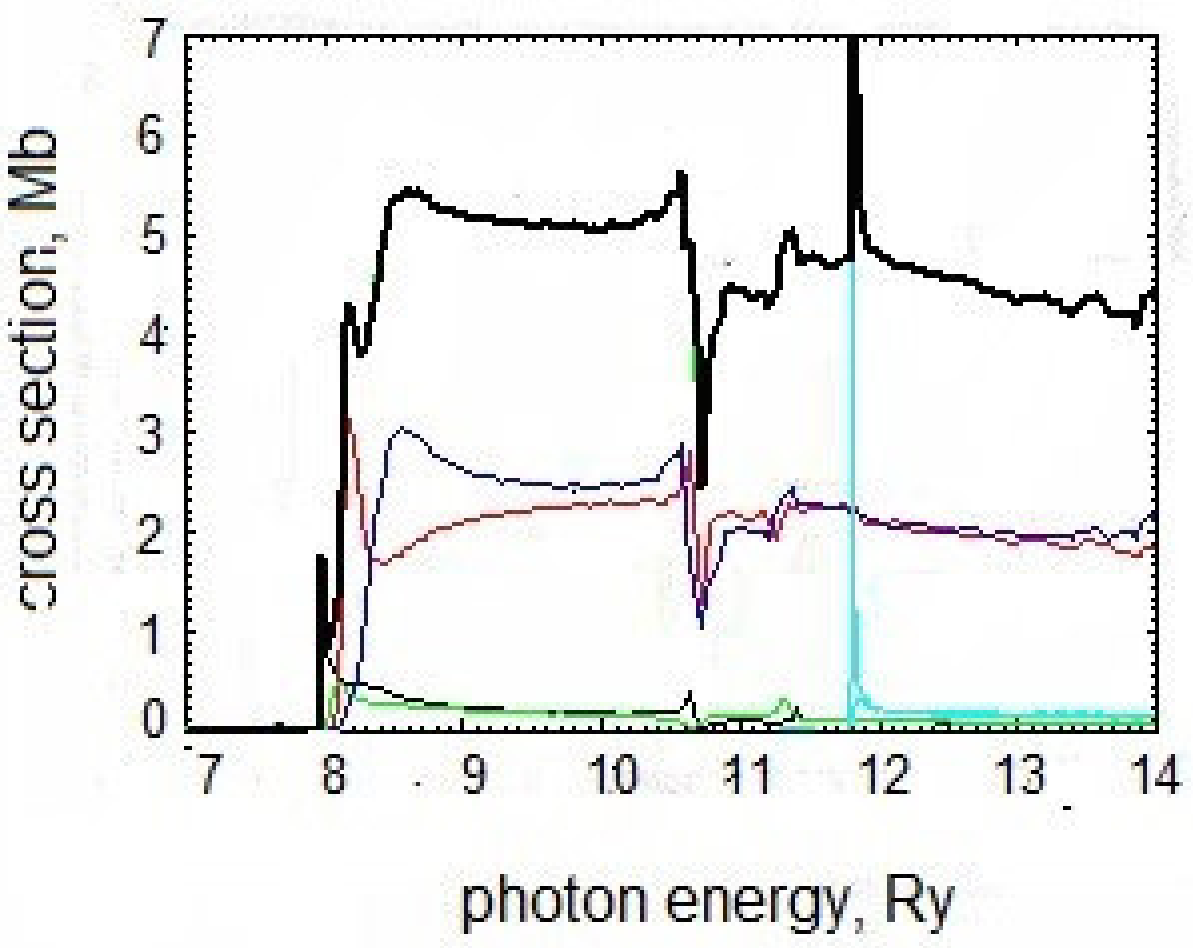